\documentclass{article}
\usepackage{amssymb}

\usepackage{amsmath}
\usepackage{graphicx}

\def\be{\begin{equation}}
\def\ee{\end{equation}}

\def\bea{\begin{eqnarray}}
\def\eea{\end{eqnarray}}

\begin{document}

\title{On Problems of the Lagrangian Quantization of $W_{3}$-gravity}
\date{}
\author{B. Geyer$^{a),b)}$\thanks{%
E-mail: geyer@itp.uni-leipzig.de}, D.M. Gitman$^{b)}$\thanks{%
E-mail: gitman@dfn.if.usp.br}, P.M. Lavrov$^{c)}$\thanks{%
E-mail: lavrov@tspu.edu.ru} and P.Yu. Moshin$^{b),c)}$\thanks{%
E-mail: moshin@dfn.if.usp.br} \\
\\
$^{a)}$Center of Theoretical Studies, Leipzig University,\\
Augustusplatz 10/11, D-04109 Leipzig, Germany\\
\\
$^{b)}$Instituto de F{\'{\i}}sica, Universidade de S\~{a}o Paulo,\\
Caixa Postal 66318-CEP, 05315-970 S\~{a}o Paulo, S.P., Brazil\\
\\
$^{c)}$Tomsk State Pedagogical University, 634041 Tomsk, Russia}
\maketitle

\begin{abstract}
We consider the two-dimensional model of $W_{3}$-gravity within Lagrangian
quantization methods for general gauge theories. We use the
Batalin--Vilkovisky formalism to study the arbitrariness in the realization
of the gauge algebra. We obtain a one-parametric non-analytic extension of
the gauge algebra, and a corresponding solution of the classical master
equation, related via an anticanonical transformation to a solution
corresponding to an analytic realization. We investigate the possibility of
closed solutions of the classical master equation in the Sp(2)-covariant
formalism and show that such solutions do not exist in the approximation up
to the third order in ghost and auxiliary fields.
\end{abstract}

\section{Introduction}

The two-dimensional model of $W_{3}$-gravity proposed in \cite{Hu} is an
example of an irreducible gauge theory with an open algebra of gauge
transformations. The covariant quantization of this model in the
Batalin--Vilkovisky (BV) formalism \cite{BV} was considered in \cite
{Hu,VVanP,DeJV}. In particular, in \cite{VVanP,DeJV} a closed solution of
the classical master equation (CME) was obtained. In \cite{VVanP}, the
arbitrariness of solutions was discussed in terms of anticanonical
transformations \cite{antican}, which effectively describe the arbitrariness
in the gauge algebra.

In this paper, we continue to investigate the aspects of Lagrangian
quantization of the model \cite{Hu}, namely, we extend the study of the
gauge algebra, using solutions of CME in the BV formalism, and analyze the
possibility of finding a closed solution of CME in the Sp(2)-covariant
quantization \cite{blt}.

Our interest in the Lagrangian quantization of the model is due to its
peculiarities at the classical level. On the one hand, the Hamiltonian
structure of the model does not conform to the assumptions \cite{GiTyu} that
guarantee the applicability of some general statements established in the
theory of constraint systems. On the other hand, the structure of extremals
and gauge generators in the Lagrangian formulation leads to a freedom in the
definition of the structure functions that admits representations with
non-analytic structure functions.

Since the model admits a closed solution of CME, this allows one to study
the arbitrariness in the structure functions using anticanonical
transformations of solutions. We obtain a realization of the gauge algebra
depending on a free parameter and corresponding, for non-vanishing values of
this parameter, to non-analytic realizations of the structure functions. The
solution of CME constructed in \cite{DeJV} corresponds to an analytic
representation of the gauge algebra. We obtain a closed solution of CME in
the case of non-analytic structure functions. This solution is related to 
\cite{DeJV} via an anticanonical transformation and contains structure
functions of the third level. We notice that the arbitrariness in the gauge
algebra found in \cite{VVanP} is the unique arbitrariness, with the given
set of generators, that preserves analyticity and ensures the absence of
structure functions beyond the second level.

The fact that $W_{3}$-gravity has a closed solution \cite{DeJV} of CME in
the BV formalism also explains our interest in the Sp(2)-covariant
quantization of this model. Note that the solution \cite{DeJV} is based on a
choice of the gauge algebra where the openness is described by a single
structure function, being also linear in the fields. The choice \cite{DeJV}
provides the simplest realization of the gauge algebra in the analytic class 
\cite{VVanP}, for which one could attempt to find a closed solution of CME
in the Sp(2)-covariant formalism despite the considerably more complicated
structure of the corresponding approximated solutions \cite{lmr}.

Using the general results for approximated solutions \cite{lmr} in the
Sp(2)-covariant formalism and the realization of the gauge algebra \cite
{DeJV}, we show that a closed solution of CME for $W_{3}$-gravity does not
exist up to the third order in the ghost and auxiliary variables. The
example of $W_{3}$-gravity shows that for theories with an open algebra the
existence of a closed solution of CME in the BV formalism up to the second
order does not guarantee the existence of a closed solution of CME in the
Sp(2)-covariant formalism up to the third order, in contrast to the case of
irreducible theories with a closed algebra.

The paper is organized as follows. In Sect. 2, we discuss the peculiarities
of the model in the Lagrangian and Hamiltonian formulations. In Sect. 3, we
solve the gauge algebra at the second level, present solutions of CME
corresponding to different choices of the gauge algebra and discuss the
results from the viewpoint of anticanonical transformations.\ In Sect. 4, we
consider the model within the Sp(2)-covariant formalism. In the Appendix, we
present the details of the Sp(2)-covariant calculations.

\section{Peculiarities of the Lagrangian and Hamiltonian formulations of the
model}

The model of $W_{3}$-gravity \cite{Hu,VVanP,DeJV} is described by the
classical action of the form

\begin{equation}
S_{0}=\int \mathcal{L\,}d^{2}x\,,\;\;\;\mathcal{L}=\frac{1}{2}\phi ^{\prime }%
\dot{\phi}-\frac{1}{2}h\phi ^{\prime \,2}-\frac{1}{3}B\phi ^{\prime \,3}.
\label{w3act}
\end{equation}
Here, the bosonic fields $(\phi ,h,B)$ are defined on a two-dimensional
space with coordinates $x=(x^{+},x^{-})$,\ and the following notation is
used: 
\begin{equation*}
\phi ^{\prime }\equiv \partial \phi =\frac{\partial \phi }{\partial x^{+}}%
\,,\;\;\;\dot{\phi}\equiv \bar{\partial}\phi =\frac{\partial \phi }{\partial
x^{-}}\,,\;\;\;d^{2}x=dx^{+}dx^{-}\,.
\end{equation*}

The equations of motion read 
\begin{align}
\frac{\delta S_{0}}{\delta \phi }& =(-\dot{\phi}+h\phi ^{\prime }+B\phi
^{\prime \,2})^{\prime }=0,  \notag \\
\frac{\delta S_{0}}{\delta h}& =-\frac{1}{2}\phi ^{\prime \,2}=0,\quad \frac{%
\delta S}{\delta B}=-\frac{1}{3}\phi ^{\prime \,3}=0,  \label{eq_m}
\end{align}
and obviously imply 
\begin{equation*}
\phi ^{\prime }=0.
\end{equation*}

The action (\ref{w3act}) is invariant under gauge transformations of the
form \cite{Hu,DeJV} 
\begin{align}
\delta \phi & =\phi ^{\prime }\epsilon +\phi ^{\prime \,\,2}\lambda ,  \notag
\\
\delta h& =\dot{\epsilon}-h\epsilon ^{\prime }+h^{\prime }\epsilon +\phi
^{\prime \,2}(B^{\prime }\lambda -B\lambda ^{\prime }),  \notag \\
\delta B& =B^{\prime }\epsilon -2B\epsilon ^{\prime }+\dot{\lambda}-h\lambda
^{\prime }+2h^{\prime }\lambda  \label{g_tr}
\end{align}
with the bosonic parameters $\xi ^{\alpha }=(\epsilon ,\lambda )$. Denoting $%
A^{i}=(\phi ,h,B)$,\ and $\delta A^{i}=R_{\alpha }^{i}(A)\xi ^{\alpha }$, we
have the following identification of the gauge generators $R_{\alpha }^{i}$: 
\begin{align}
R_{\alpha }^{\phi }& =\left( \phi ^{\prime },\;\;\phi ^{\prime \,2}\right) ,
\notag \\
R_{\alpha }^{h}& =\left( \bar{\partial}-h\partial +h^{\prime },\;\;\phi
^{\prime \,2}(B^{\prime }-B\partial )\right) ,  \notag \\
R_{\alpha }^{B}& =\left( B^{\prime }-2B\partial ,\;\;\bar{\partial}%
-h\partial +2h^{\prime }\right) .  \label{g_gen}
\end{align}

The generators $(R_{\alpha }^{h},R_{\alpha }^{B})$ do not vanish on shell,
and therefore are non-trivial. The generator $R_{\alpha }^{\phi }$ vanishes
on shell. However, it cannot be presented as an action of a local operator
on the extremals. Thus, we encounter a non-typical case. As a consequence,
the gauge algebra of the model may contain non-analytic structure functions.
In the following section, we shall see that this is indeed the fact.

It should be noted that the properties of the complete theory defined by the
action $S_{0}$ are essentially different from the properties of its
quadratic approximation defined by the action $S_{0}^{(2)}=\left.
S_{0}\right| _{h=B=0}\,$. In particular, the quadratic theory is not a gauge
one. The obvious reparametrization invariance of the latter model belongs to
a wider class of symmetries, having the form 
\begin{equation}
\delta \phi =\sum_{s=1}(k_{s}\phi ^{\prime }{}^{\,s-1}+\tilde{k}_{s}\dot{\phi%
}^{\,s-1}),\;\;\;k_{s}=k_{s}(x^{+}),\;\;\;\tilde{k}_{s}=\tilde{k}_{s}(x^{-}).
\label{semirigid}
\end{equation}
Particular cases of (\ref{semirigid}), namely, $s=2$ and $\tilde{k}_{s}=0$,
have been discussed in \cite{pope}. The presence of restrictions $\dot{k}%
_{s}=\tilde{k}_{s}^{\prime }=0$ on the parameters does not allow one to
treat (\ref{semirigid}) as gauge transformations. Another argument in favour
of this interpretation can be found in the Hamiltonian formalism, which does
not have first-class constraints, and thus non-trivial gauge invariance is
absent.

One may expect that the properties of the complete theory and its quadratic
approximation should also be essentially different in the Hamiltonian
formulation. Namely, the constraint structure of the complete theory and
that of its quadratic approximation must be different. To analyze this
problem in more detail, let us construct the Hamiltonian formulation of the
theory under consideration. We select $x^{-}$ to be the time variable. In
this case, the corresponding Hessian matrix has a constant rank on shell.
There are three primary constraints: 
\begin{align*}
p_{\phi }& =\frac{\partial \mathcal{L}}{\partial \dot{\phi}}=\frac{1}{2}\phi
^{\prime }\Longrightarrow \Phi _{1}^{\left( 1\right) }=p_{\phi }-\phi
^{\prime }/2=0, \\
p_{h}& =\frac{\partial \mathcal{L}}{\partial \dot{h}}=0\Longrightarrow \Phi
_{2}^{\left( 1\right) }=p_{h}=0, \\
p_{B}& =\frac{\partial \mathcal{L}}{\partial \dot{B}}=0\Longrightarrow \Phi
_{3}^{\left( 1\right) }=p_{B}=0.
\end{align*}
The total Hamiltonian reads 
\begin{equation*}
H^{\left( 1\right) }=H+\lambda _{i}\Phi _{i}^{\left( 1\right)
}\,,\;\;\;H=\int \left( \frac{1}{2}h\phi ^{\prime \,\,2}+\frac{1}{3}B\phi
^{\prime \,3}\right) dx^{+}\,,
\end{equation*}
The consistency conditions for the primary constraints imply the secondary
constraint $\Phi ^{\left( 2\right) }=\phi ^{\prime }=0$\ and define one of $%
\lambda $'s, namely,\ $\lambda _{1}^{\prime }=0.$ No more secondary
constraints appear, and $\lambda _{2}$, $\lambda _{3}$ remain undetermined.
An equivalent complete set of constraints can be written as 
\begin{align}
\Phi _{2}^{\left( 1\right) }& =p_{h}\,,\quad \Phi _{3}^{\left( 1\right)
}=p_{B}\,,  \label{1st_class} \\
\tilde{\Phi}_{1}^{\left( 1\right) }& =p_{\phi }\,,\quad \Phi ^{\left(
2\right) }=\phi ^{\prime }\,.  \label{2nd_class}
\end{align}
Here, (\ref{1st_class}) are first-class constraints, while (\ref{2nd_class})
are second-class constraints.

One can easily see that in the Hamiltonian formulation of the quadratic
theory with the action $S_{0}^{(2)}$ there exists only one primary
constraint $\Phi ^{\left( 1\right) }\equiv p_{\phi }-\phi ^{\prime }/2=0$,
with $H^{\left( 1\right) }=\lambda \Phi ^{\left( 1\right) }$. The constraint 
$\Phi ^{\left( 1\right) }$ is a second-class one\footnote{%
Here and elsewhere, we use the notation $\delta ^{\prime }(x)=\partial
\delta (x)$.} 
\begin{equation*}
\left\{ \Phi ^{\left( 1\right) }(x_{1}^{+}),\Phi ^{\left( 1\right)
}(x_{2}^{+})\right\} =-\delta ^{\prime }(x_{1}^{+}-x_{2}^{+}).
\end{equation*}

It is obvious that the constraints of the quadratic and complete theories
have a different structure: the constraints of the complete theory are not
the constraints of the quadratic theory with non-linear corrections to them.
This peculiarity of the model (\ref{w3act}) does not conform to the
assumptions \cite{GiTyu} which guarantee the applicability of general
statements established in the theory of constraint systems. This means that
the Hamiltonian quantization of the given model may encounter difficulties.
Note that the choice of the time variable as $x^{+}$ leads to a Hessian
matrix whose rank is not constant in the vicinity of the zero point, being a
natural point of consideration in the perturbation theory.

It should be noted that the model with the action (\ref{w3act}) and
generators (\ref{g_gen}) satisfies the conditions that guarantee the
applicability of Lagrangian quantization \cite{BV}. Indeed, there exists a
stationary point $(\phi ^{\prime }=0)$ in whose neighborhood the action and
gauge generators are analytic. Besides, at the stationary point there hold
the rank conditions 
\begin{equation}
\mathrm{rank\,}\left\| S_{0,ij}(A)\right\| _{A=A_{0}}=n-m,\;\;\;\mathrm{%
rank\,}\left\| R_{\alpha }^{i}(A)\right\| _{A=A_{0}}=m,  \label{rank}
\end{equation}
where $A_{0}^{i}$ denotes the stationary point; rank is understood with
respect to the discrete indices $i=1,...,n$, $\alpha =1,...,m$; and $%
S_{0,ij}(A)$ is given by 
\begin{equation*}
S_{0,ij}(A)\equiv \frac{\delta ^{2}S_{0}(A)}{\delta A^{i}\delta A^{j}}.
\end{equation*}
Calculating the matrices $\left\| S_{0,ij}(A)\right\| _{A=A_{0}}$ and $%
\left\| R_{\alpha }^{i}(A)\right\| _{A=A_{0}}$ in the model (\ref{w3act}), (%
\ref{g_gen}), we find 
\begin{equation}
\left\| S_{0,ij}(A)\right\| _{A=A_{0}}=\left( 
\begin{array}{ccc}
-\partial \bar{\partial}+\partial (h\partial ) & 0 & 0 \\ 
0 & 0 & 0 \\ 
0 & 0 & 0
\end{array}
\right) \delta (x_{1}-x_{2})  \label{hess}
\end{equation}
and 
\begin{equation}
\left\| R_{\alpha }^{i}(A)\right\| _{A=A_{0}}=\left( 
\begin{array}{ccc}
0 & \bar{\partial}-h\partial +h^{\prime } & B^{\prime }-2B\partial \\ 
0 & 0 & \bar{\partial}-h\partial +2h^{\prime }
\end{array}
\right) ,  \label{r}
\end{equation}
where the rows and columns of (\ref{hess}) are labeled by $(\phi ,h,B)$,
while the rows and columns of (\ref{r}) are labeled by $(\epsilon ,\lambda )$
and $(\phi ,h,B)$, respectively. 
%
From (\ref{hess}) and (\ref{r}), it follows that the conditions (\ref{hess})
and (\ref{r}) are satisfied, with $n=3$, $m=2$. The fulfillment of the rank
conditions (\ref{rank}) implies that the set of gauge generators (\ref{g_gen}%
) is complete and linearly independent \cite{BV}. Thus, the model of $W_{3}$%
-gravity can be quantized using Lagrangian methods for theories with
linearly independent (irreducible) generators.

\section{Extended gauge algebra of the model}

%
In this section, we shall consider the gauge algebra of the model with the
action (\ref{w3act}) and the gauge generators (\ref{g_gen}). The gauge
generators determine the first level of the gauge algebra, given by the
corresponding Noether identities $S_{0,i}(A)R_{\alpha }^{i}(A)=0$. At the
second level, the gauge algebra coincides with the algebra of gauge
generators with respect to their commutator, given by 
\begin{equation*}
\lbrack \delta _{1},\delta _{2}]A^{i}=\left[ R_{\alpha ,j}^{i}(A)R_{\beta
}^{j}(A)-R_{\beta ,j}^{i}(A)R_{\alpha }^{j}(A)\right] \xi _{1}^{\beta }\xi
_{2}^{\alpha },
\end{equation*}
where $\delta _{1}$, $\delta _{2}$ correspond to gauge transformations with
parameters $\xi _{1}^{\beta }$, $\xi _{2}^{\alpha }$. Then, using the
explicit form of gauge transformations (\ref{g_tr}) and identifying $\xi
_{1}^{\beta }=(\epsilon _{1},\lambda _{1})$, $\xi _{2}^{\alpha }=(\epsilon
_{2},\lambda _{2})$, we get 
\begin{align}
\lbrack \delta _{1},\delta _{2}]\phi & =-\phi ^{\prime }\epsilon
_{(1,2)}-\phi ^{\prime \,2}\left[ (\epsilon \lambda )_{(1,2)}-(\epsilon
\lambda )_{(2,1)}\right] -2\phi ^{\prime \,3}\lambda _{(1,2)},  \notag \\
\lbrack \delta _{1},\delta _{2}]h& =-(\bar{\partial}-h\partial +h^{\prime
})\epsilon _{(1,2)}-\phi ^{\prime \,2}(B^{\prime }-B\partial )\left[
(\epsilon \lambda )_{(1,2)}-(\epsilon \lambda )_{(2,1)}\right]  \notag \\
& -\phi ^{\prime \,2}(\bar{\partial}-h\partial +3h^{\prime }+2\phi ^{\prime
}B^{\prime }+4\phi ^{\prime \prime }B)\lambda _{(1,2)},  \notag \\
\lbrack \delta _{1},\delta _{2}]B& =-(B^{\prime }-2B\partial )\epsilon
_{(1,2)}-(\bar{\partial}-h\partial +2h^{\prime })[(\epsilon \lambda
)_{(1,2)}-(\epsilon \lambda )_{(2,1)}]  \notag \\
& -(\phi ^{\prime \,2}B^{\prime }-4\phi ^{\prime }\phi ^{\prime \prime
}B-2\phi ^{\prime \,2}B\partial )\lambda _{(1,2)},  \label{w3alg1}
\end{align}
where we have used the notation 
\begin{equation}
\epsilon _{(1,2)}=\epsilon _{1}\epsilon _{2}^{\prime }-\epsilon _{1}^{\prime
}\epsilon _{2},\quad (\epsilon \lambda )_{(1,2)}=\epsilon _{1}\lambda
_{2}^{\prime }-2\epsilon _{1}^{\prime }\lambda _{2},\quad \lambda
_{(1,2)}=\lambda _{1}\lambda _{2}^{\prime }-\lambda _{1}^{\prime }\lambda
_{2}.  \label{param2}
\end{equation}

To analyze the relations (\ref{w3alg1}), we remind that in the bosonic case
of a gauge theory with a complete set of generators the commutator of gauge
generators has the general form \cite{BV} 
\begin{equation}
R_{\alpha,j}^{i}(A)R_{\beta}^{j}(A)-R_{\beta,j}^{i}(A)R_{\alpha}^{j}(A)=-R_{%
\gamma}^{i}(A)F_{\alpha\beta}^{\gamma}(A)-S_{0,j}(A)M_{\alpha \beta}^{ij}(A),
\label{GAGGT}
\end{equation}
where $F_{\alpha\beta}^{\gamma}(A)$ and $M_{\alpha\beta}^{ij}(A)$ are
structure functions, generally depending on the fields $A^{i}$ and
possessing the antisymmetry properties 
\begin{equation*}
F_{\alpha\beta}^{\gamma}(A)=-F_{\beta\alpha}^{\gamma}(A),\;\;\;M_{\alpha%
\beta }^{ij}(A)=-M_{\alpha\beta}^{ji}(A)=-M_{\beta\alpha}^{ij}(A).
\end{equation*}
The set of structure functions $F_{\alpha\beta}^{\gamma}$ and $%
M_{\alpha\beta }^{ij}$ defines the gauge algebra at the second level. 
%
If $M_{\alpha\beta}^{ij}(A)=0$, the gauge algebra is closed. If $M_{\alpha
\beta}^{ij}(A)\neq0$, it is open.

Taking into account the general structure (\ref{GAGGT}) of the algebra of
generators and the explicit form (\ref{g_gen}) of gauge generators in the
model of $W_{3}$-gravity, one gets from (\ref{w3alg1}) the following
structure functions $F_{11}^{1}$, $F_{21}^{2}$,$\,F_{12}^{2}$: 
\begin{equation}
\epsilon _{(1,2)}=F_{11}^{1}\epsilon _{1}\epsilon _{2},\quad (\epsilon
\lambda )_{(1,2)}=F_{21}^{2}\epsilon _{1}\lambda _{2},\quad (\epsilon
\lambda )_{(2,1)}=-F_{12}^{2}\epsilon _{2}\lambda _{1},  \label{param3}
\end{equation}
namely, 
\begin{align}
F_{11}^{1}& \equiv F_{\epsilon (y_{1})\epsilon (y_{2})}^{\epsilon
(x)}=\delta (x-y_{2})\delta ^{\prime }(x-y_{1})-\delta (x-y_{1})\delta
^{\prime }(x-y_{2}),  \notag \\
F_{21}^{2}& \equiv F_{\lambda (y_{1})\epsilon (y_{2})}^{\lambda (x)}=\delta
(x-y_{2})\delta ^{\prime }(x-y_{1})-2\delta (x-y_{1})\delta ^{\prime
}(x-y_{2}),  \notag \\
F_{12}^{2}& \equiv F_{\epsilon (y_{1})\lambda (y_{2})}^{\lambda (x)}=-\left[
\delta (x-y_{1})\delta ^{\prime }(x-y_{2})-2\delta (x-y_{2})\delta ^{\prime
}(x-y_{1})\right] .  \label{w3SF1}
\end{align}
Obviously, these structure functions do not depend on the fields.

The structure functions related to the terms of the gauge algebra containing 
$\lambda _{(1,2)}$ are not so easy to determine, since, besides $R_{\alpha
}^{i}$, they also contain the structure functions $M_{\alpha \beta }^{ij}$.
By virtue of the relation (\ref{w3alg1}) for $[\delta _{1},\delta _{2}]\phi $%
, it is natural to suggest the following Ansatz for the remainder: 
\begin{equation}
2\phi ^{\prime \,3}\lambda _{(1,2)}=\left( R_{1}^{\phi
}F_{22}^{1}+R_{2}^{\phi }F_{22}^{2}+\frac{\delta S_{0}}{\delta h}%
M_{22}^{\phi h}+\frac{\delta S_{0}}{\delta B}M_{22}^{\phi B}\right) \lambda
_{1}\lambda _{2}.  \label{ansatz}
\end{equation}
According to (\ref{eq_m}) and (\ref{g_gen}), we can parameterize $F_{22}^{1}$%
, $F_{22}^{2}$, $M_{22}^{\phi h}$,$\ M_{22}^{\phi B}$ as follows: 
\begin{align}
F_{22}^{1}\lambda _{1}\lambda _{2}& =\alpha _{1}\phi ^{\prime \,2}\lambda
_{(1,2)},\qquad F_{22}^{2}\lambda _{1}\lambda _{2}=\alpha _{2}\phi ^{\prime
}\lambda _{(1,2)},  \notag \\
M_{22}^{\phi h}\lambda _{1}\lambda _{2}& =2\beta _{1}\phi ^{\prime }\lambda
_{(1,2)},\qquad M_{22}^{\phi B}\lambda _{1}\lambda _{2}=3\beta _{2}\lambda
_{(1,2)},  \label{param}
\end{align}
where the constant parameters $\alpha _{1}$, $\alpha _{2}$, $\beta _{1}$, $%
\beta _{2}$ must satisfy the relation 
\begin{equation}
\alpha _{1}+\alpha _{2}-\beta _{1}-\beta _{2}=2.  \label{rel}
\end{equation}
Returning with these results to the remainder of $[\delta _{1},\delta _{2}]h$
and $[\delta _{1},\delta _{2}]B$, we shall seek the structure functions in
the form 
\begin{align}
& \phi ^{\prime \,2}(\bar{\partial}-h\partial +3h^{\prime }+2\phi ^{\prime
}B^{\prime }+4\phi ^{\prime \prime }B)\lambda _{(1,2)}=\left(
R_{1}^{h}F_{22}^{1}+R_{2}^{h}F_{22}^{2}+\frac{\delta S_{0}}{\delta \phi }%
M_{22}^{h\phi }+\frac{\delta S_{0}}{\delta B}M_{22}^{hB}\right) \lambda
_{1}\lambda _{2},  \notag \\
& (\phi ^{\prime \,2}B^{\prime }-4\phi ^{\prime }\phi ^{\prime \prime
}B-2\phi ^{\prime \,2}B\partial )\lambda _{(1,2)}=\left(
R_{1}^{B}F_{22}^{1}+R_{2}^{B}F_{22}^{2}+\frac{\delta S_{0}}{\delta \phi }%
M_{22}^{B\phi }+\frac{\delta S_{0}}{\delta h}M_{22}^{Bh}\right) \lambda
_{1}\lambda _{2}.  \label{ansatz2}
\end{align}
Within this Ansatz, the only structure coefficient left to determine is $%
M_{22}^{hB}$. Let us represent it by a certain operator $M(A^{i},\partial ,{%
\bar{\partial}})$, as follows: 
\begin{equation*}
M_{22}^{hB}\lambda _{1}\lambda _{2}=M\lambda _{(1,2)}\,.
\end{equation*}
Then, using the identification (\ref{g_gen}) and the condition (\ref{rel}),
we find the following relations, which explicitly realize the second-level
gauge algebra of $W_{3}$-gravity: 
\begin{equation*}
\alpha _{1}=1,\quad \alpha _{2}=0,\quad \beta _{1}+\beta _{2}=-1,\;\;\phi
^{\prime 2}M=6\beta _{2}\frac{\delta S_{0}}{\delta \phi }.
\end{equation*}
Thus, the Ansatz (\ref{ansatz}), (\ref{ansatz2}) determines the algebra with
accuracy up to a free parameter, $\beta _{2}\equiv \beta $. Together with (%
\ref{w3SF1}), we obtain the remaining non-vanishing structure functions of
the second level: 
\begin{align}
F_{22}^{1}& =\phi ^{\prime \,2}\left[ \delta (x-y_{2})\delta ^{\prime
}(x-y_{1})-\delta (x-y_{1})\delta ^{\prime }(x-y_{2})\right] ,  \notag \\
M_{22}^{\phi h}& =-2(1+\beta )\phi ^{\prime }\delta (x-y)\left[ \delta
(y-y_{2})\delta ^{\prime }(y-y_{1})-\delta (y-y_{1})\delta ^{\prime
}(y-y_{2})\right] ,  \notag \\
M_{22}^{\phi B}& =3\beta \delta (x-y)\left[ \delta (y-y_{2})\delta ^{\prime
}(y-y_{1})-\delta (y-y_{1})\delta ^{\prime }(y-y_{2})\right] ,  \notag \\
M_{22}^{hB}& =6\beta \phi ^{\prime }{}^{\,-2}\frac{\delta S_{0}}{\delta \phi 
}\delta (x-y)\left[ \delta (y-y_{2})\delta ^{\prime }(y-y_{1})-\delta
(y-y_{1})\delta ^{\prime }(y-y_{2})\right] .  \label{w3fm}
\end{align}

If $\beta \neq 0$, then we have a realization of the algebra with a
coefficient $M_{22}^{hB}$ non-analytic at the stationary point of $S_{0}$.
Under the requirement of analyticity $(\beta =0)$ we obtain a realization of
the gauge algebra of $W_{3}$-gravity with the gauge generators $R_{\alpha
}^{i}$ (\ref{g_gen}) and the following non-vanishing structure functions $%
F_{\alpha \beta }^{\gamma }$, $M_{\alpha \beta }^{ij}$: 
\begin{align}
F_{11}^{1}& =\delta (x-y_{2})\delta ^{\prime }(x-y_{1})-\delta
(x-y_{1})\delta ^{\prime }(x-y_{2}),  \notag \\
F_{22}^{1}& =\phi ^{\prime \,2}\left[ \delta (x-y_{2})\delta ^{\prime
}(x-y_{1})-\delta (x-y_{1})\delta ^{\prime }(x-y_{2})\right] ,  \notag \\
F_{21}^{2}& =\delta (x-y_{2})\delta ^{\prime }(x-y_{1})-2\delta
(x-y_{1})\delta ^{\prime }(x-y_{2})  \label{w3F}
\end{align}
and 
\begin{equation}
M_{22}^{\phi h}=2\phi ^{\prime }\delta (x-y)[\delta (y-y_{1})\delta ^{\prime
}(y-y_{2})-\delta (y-y_{2})\delta ^{\prime }(y-y_{1}).  \label{w3M}
\end{equation}
This is the realization of the gauge algebra of $W_{3}$-gravity which was
used in \cite{DeJV} to construct a solution of the classical master
equation. The realization (\ref{w3fm}) provides a non-analytic extension of (%
\ref{w3F}), (\ref{w3M}). We can see that the model of $W_{3}$-gravity
belongs to the class of irreducible gauge theories with an open algebra of
gauge generators.

To complete the definition of the gauge algebra, it is necessary to add also
the set of structure functions at higher levels \cite{BV}. For irreducible
theories, this set can be derived by using: (i) the Jacobi identity for the
commutator of gauge transformations, (ii) the conditions of completeness and
irreducibility, and (iii) the previous gauge relations, such as the Noether
identities or relations (\ref{GAGGT}). In general, the gauge algebra
consists of an infinite set of structure functions, which define an infinite
number of structure relations.

For the purpose of this paper, it is sufficient to know the structure of the
gauge algebra up to the third level. Let us consider the Jacobi identity for
the gauge transformations 
\begin{equation*}
\lbrack \delta _{1},[\delta _{2},\delta _{3}]]A^{i}+\mathrm{cycl.perm.}%
(1,2,3)\equiv 0.
\end{equation*}
Then, using differential consequences of the Noether identities $%
S_{0,i}(A)R_{\alpha }^{i}(A)=0$, we obtain 
\begin{equation}
\left( R_{\gamma }^{i}D_{\alpha \beta \delta }^{\gamma }+S_{0,k}Z_{\alpha
\beta \delta }^{ik}\right) \xi _{1}^{\delta }\xi _{2}^{\beta }\xi
_{3}^{\alpha }+\mathrm{cycl.perm.}(1,2,3)=0,  \label{GAnR}
\end{equation}
with the following abbreviations: 
\begin{align}
D_{\alpha \beta \delta }^{\gamma }& \equiv \left( F_{\alpha \sigma }^{\gamma
}F_{\beta \delta }^{\sigma }+F_{\alpha \beta ,i}^{\gamma }R_{\delta
}^{i}\right) +\mathrm{cycl.perm.}(\alpha ,\beta ,\delta ),  \notag \\
Z_{\alpha \beta \delta }^{ik}& \equiv \left( M_{\alpha \sigma }^{ik}F_{\beta
\delta }^{\sigma }+M_{\alpha \beta ,j}^{ik}R_{\delta }^{j}-R_{\alpha
,j}^{k}M_{\beta \delta }^{ij}+R_{\alpha ,j}^{i}M_{\beta \delta }^{kj}\right)
+\mathrm{cycl.perm.}(\alpha ,\beta ,\delta ),  \notag
\end{align}
where $D_{\alpha \beta \delta }^{\gamma }$ and $Z_{\alpha \beta \delta
}^{ik} $ are antisymmetric in the indices $(\alpha ,\beta ,\delta )$ and $%
(i,k)$.

By virtue of the completeness and linear independence of the generators $%
R_{\alpha}^{i}$, the relation (\ref{GAnR}) is solved by \cite{BV} 
\begin{equation}
D_{\alpha\beta\delta}^{\gamma}=S_{0,k}Q_{\alpha\beta\delta}^{\gamma
k},\;\;\;Z_{\alpha\beta\delta}^{ik}+R_{\gamma}^{i}Q_{\alpha\beta\delta
}^{\gamma k}-R_{\gamma}^{k}Q_{\alpha\beta\delta}^{\gamma
i}=S_{0,j}M_{\alpha\beta\delta}^{ikj},  \label{GAnR1}
\end{equation}
where $Q_{\alpha\beta\delta}^{\gamma k}$ and $M_{\alpha\beta\delta}^{ikj}$
are structure functions, antisymmetric in the indices $(\alpha,\beta,\delta)$
and $(i,j,k)$. For irreducible theories, the functions $Q_{\alpha\beta\delta
}^{\gamma k}$ and $M_{\alpha\beta\delta}^{ikj}$ define the structure of the
gauge algebra at the third level.

To analyze the structure functions of $W_{3}$-gravity beyond the second
level, it is convenient to apply the BV formalism \cite{BV}. Within this
formalism, all structure relations can be collected into a solution of the
classical master equation. This equation is formulated for the bosonic
extended action $S=S(\phi ,\phi ^{\ast })$. For irreducible theories, the
action depends on the minimal set of classical and ghost fields $\phi
^{A}=(A^{i},C^{\alpha })$, $\varepsilon (C^{\alpha })=\varepsilon (\xi
^{\alpha })+1$, and the corresponding antifields $\phi _{A}^{\ast
}=(A_{i}^{\ast },C_{\alpha }^{\ast })$,$\;\varepsilon (\phi _{A}^{\ast
})=\varepsilon (\phi ^{A})+1$, with the following distribution of the ghost
number: 
\begin{equation*}
\mathrm{gh}(A^{i})=0,\quad \mathrm{gh}(C^{\alpha })=1,\quad \mathrm{gh}(\phi
_{A}^{\ast })=-1-\mathrm{gh}(\phi ^{A}).
\end{equation*}
The classical master equation for the gauge algebra is defined by 
\begin{equation}
\frac{\delta S}{\delta \phi ^{A}}\frac{\delta S}{\delta \phi _{A}^{\ast }}=0,
\label{CME}
\end{equation}
and is subject to the boundary condition 
\begin{equation*}
S|_{\phi ^{\ast }=0}=S_{0}(A).
\end{equation*}
A solution $S=S(\phi ,\phi ^{\ast })$ can be sought as a Taylor series in
the ghost fields $C^{\alpha }$, 
\begin{equation*}
S=S_{0}(A)+\sum_{n=1}S_{n},\quad S_{n}\sim (C)^{n},\quad \varepsilon
(S_{n})=0,\;\;\mathrm{gh}(S_{n})=0,
\end{equation*}
with the following result (see, e.g., \cite{GPS}), considered in the bosonic
case $\varepsilon (A^{i})=\varepsilon (\xi ^{\alpha })=0$, with accuracy up
to the third order: 
\begin{align}
S(\phi ,\phi ^{\ast })& =S_{0}(A)+A_{i}^{\ast }R_{\alpha }^{i}C^{\alpha }-%
\frac{1}{2}\left( C_{\gamma }^{\ast }F_{\alpha \beta }^{\gamma }-\frac{1}{2}%
A_{i}^{\ast }A_{j}^{\ast }M_{\alpha \beta }^{ij}\right) C^{\beta }C^{\alpha }
\notag \\
& -\frac{1}{2}\left( C_{\delta }^{\ast }A_{i}^{\ast }Q_{\alpha \beta \gamma
}^{\delta i}-\frac{1}{6}A_{i}^{\ast }A_{j}^{\ast }A_{k}^{\ast }M_{\alpha
\beta \gamma }^{ijk}\right) C^{\gamma }C^{\beta }C^{\alpha }+\cdots \,,
\label{AcIr2BV}
\end{align}
where $F_{\alpha \beta }^{\gamma }$, $M_{\alpha \beta }^{ij}$ and $Q_{\alpha
\beta \gamma }^{\delta i}$, $M_{\alpha \beta \gamma }^{ijk}$ are the
structure functions of the gauge algebra at the second (\ref{GAGGT}) and
third (\ref{GAnR1}) levels, respectively. In the case of $W_{3}$-gravity, we
shall consider four solutions of CME, labeled by (a), (b), (c), (d).

\textbf{(a)} A closed solution of the form (\ref{AcIr2BV}) for $W_{3}$%
-gravity can be constructed \cite{VVanP,DeJV} using non-trivial structure
functions $F_{\alpha \beta }^{\gamma }$, (\ref{w3F}), and $M_{\alpha \beta
}^{ij}$, (\ref{w3M}), in the minimal sector of the classical fields $%
A^{i}=(\phi ,h,B)$ and the ghost fields $C^{\alpha }=(c,l)$: 
\begin{equation}
S=S_{0}+S_{1}+\int d^{2}x\left[ c^{\ast }\left( c^{\prime }c+\phi ^{\prime
\,\,2}l^{\prime }l\right) +l^{\ast }\left( l^{\prime }c+2c^{\prime }l\right)
+2\phi ^{\ast }h^{\ast }\phi ^{\prime }l^{\prime }l\right] ,  \label{Acw3BV}
\end{equation}
where the initial classical action $S_{0}$ is given by (\ref{w3act}), and
the action $S_{1}$ is determined by the gauge generators as 
\begin{equation*}
S_{1}=\int d^{2}x\left[ \phi ^{\ast }\left( \phi ^{\prime }c+\phi ^{\prime
\,2}l\right) +h^{\ast }\left[ \dot{c}-hc^{\prime }+h^{\prime }c+\phi
^{\prime \,2}(B^{\prime }l-Bl^{\prime })\right] +B^{\ast }\left( B^{\prime
}c-2Bc^{\prime }+\dot{l}-hl^{\prime }+2h^{\prime }l\right) \right] .
\end{equation*}
It follows from (\ref{Acw3BV}) that all structure functions of higher levels
are equal to zero if one uses the realization of the gauge algebra in the
form (\ref{g_gen}), (\ref{w3SF1}), (\ref{w3F}) and (\ref{w3M}).

\textbf{(b)} It is not difficult to construct an action (\ref{AcIr2BV}) that
corresponds to the case of the gauge algebra with non-analytic structure
functions (\ref{w3fm}). To this end, we remind that any anticanonical
transformation \cite{antican} of the field-antifield variables $\phi ^{A}$, $%
\phi _{A}^{\ast }$, determined by 
\begin{equation*}
\tilde{\phi}^{A}=\frac{\delta X(\phi ,\tilde{\phi}^{\ast })}{\delta \tilde{%
\phi}_{A}^{\ast }},\quad \phi _{A}^{\ast }=\frac{\delta X(\phi ,\tilde{\phi}%
^{\ast })}{\delta \phi ^{A}},
\end{equation*}
with the generating functional $X=X(\phi ,\;\tilde{\phi}^{\ast })$, $%
\varepsilon (X)=1$,$\;\mathrm{gh}(X)=-1$, transforms solutions of CME (\ref
{CME}) into solutions.

Making an anticanonical transformation of (\ref{Acw3BV}) with the generating
functional 
\begin{equation*}
X(\phi ,\phi ^{\ast })=E(\phi ,\phi ^{\ast })+6\beta \int d^{2}x\,\phi
^{\ast }h^{\ast }B^{\ast }\phi ^{\prime \,-2}l^{\prime }l\,,
\end{equation*}
where $E(\phi ,\phi ^{\ast })$ is the generating functional of the identical
transformation, we obtain an action with the second-level structure
functions (\ref{w3SF1}), (\ref{w3fm}),

\begin{align}
& S(\beta )\!\!=S_{0}+S_{1}+\int d^{2}x\Big[c^{\ast }(c^{\prime }c+\phi
^{\prime \,2}l^{\prime }l)+l^{\ast }(l^{\prime }c+2c^{\prime }l)  \notag \\
& +6\beta h^{\ast }B^{\ast }\phi ^{\prime \,-2}\left( \dot{\phi}^{\prime
}-h^{\prime }\phi ^{\prime }-\phi ^{\prime \prime }h-\phi ^{\prime
\,2}B^{\prime }-2\phi ^{\prime }\phi ^{\prime \prime }B\right) l^{\prime }l 
\notag \\
& -\phi ^{\ast }\left( 3\beta B^{\ast }-2(1+\beta )h^{\ast }\phi ^{\prime
}-12\beta h^{\ast }B^{\ast }\phi ^{\prime \,-2}c^{\prime }\right) l^{\prime
}l\Big].  \label{w3actsM}
\end{align}
This action, which is also a closed solution of CME, coincides with the
action (\ref{Acw3BV}) when $\beta =0$.

The realizations of the second-level gauge algebra with analytic (\ref{w3F}%
), (\ref{w3M}) and non-analytic (\ref{w3fm}) structure functions are
equivalent in the sense of the anticanonical transformation relating the
corresponding solutions (\ref{Acw3BV}) and (\ref{w3actsM}). At the same
time, from (\ref{w3actsM}) it follows that the case of the non-analytic
realization of the gauge algebra leads to a more complicated gauge
structure. Indeed, with non-vanishing $M_{122}^{\phi hB}$, one also obtains
(non-analytic) structure functions at the third level (see (\ref{GAnR1}) and
(\ref{AcIr2BV})).

\textbf{(c)} It should be noted that the discussed arbitrariness in the
choice of the gauge structure functions $F_{\alpha \beta }^{\gamma }$, $%
M_{\alpha \beta }^{ij}$ for $W_{3}$-gravity is by no means unique. Indeed,
the condition of analyticity admits a freedom in the choice of the
second-level structure functions. Namely, let us consider the action \cite
{VVanP} 
\begin{align}
\hspace{-1cm}& S(\alpha )\!\!=S_{0}+S_{1}+\int d^{2}x\Big[c^{\ast }\left(
c^{\prime }c+(1-\alpha )\phi ^{\prime \,2}l^{\prime }l\right) +l^{\ast
}\left( l^{\prime }c+2c^{\prime }l\right)  \notag \\
& +2\alpha h^{\ast }\left( \dot{h}^{\ast }-h^{\ast \prime }h-3B^{\ast
}B^{\prime }-2BB^{\ast \prime }\right) l^{\prime }l+2(1+\alpha )\phi ^{\ast
}h^{\ast }\phi ^{\prime }l^{\prime }l\Big],  \label{Acw3nBV}
\end{align}
obtained from (\ref{Acw3BV}) via anticanonical transformations with the
generating functional 
\begin{equation}
X(\phi ,\phi ^{\ast })=E(\phi ,\phi ^{\ast })-2\alpha \int d^{2}x\,h^{\ast
}c^{\ast }l^{\prime }l,  \label{w3sct}
\end{equation}
where $\alpha $ is a free parameter. The action (\ref{Acw3nBV}) satisfies
CME with the same boundary condition and gauge generators, but it
corresponds to another set of gauge structure functions $F_{\alpha \beta
}^{\gamma }$, 
\begin{align}
F_{11}^{1}& =\delta (x-y_{2})\delta ^{\prime }(x-y_{1})-\delta
(x-y_{1})\delta ^{\prime }(x-y_{2}),  \notag \\
F_{22}^{1}& =(1-\alpha )\phi ^{\prime \,2}\left[ \delta (x-y_{2})\delta
^{\prime }(x-y_{1})-\delta (x-y_{1})\delta ^{\prime }(x-y_{2})\right] , 
\notag \\
F_{21}^{2}& =\delta (x-y_{2})\delta ^{\prime }(x-y_{1})-2\delta
(x-y_{1})\delta ^{\prime }(x-y_{2}),  \label{an1}
\end{align}
and non-vanishing matrices $M_{\alpha \beta }^{ij}$, 
\begin{align}
M_{22}^{\phi h}& =2(1+\alpha )\phi ^{\prime }\delta (x-y)\left[ \delta
(y-y_{1})\delta ^{\prime }(y-y_{2})-\delta (y-y_{2})\delta ^{\prime
}(y-y_{1})\right] ,  \notag \\
M_{22}^{hh}& =2\alpha \left( {\bar{\partial}}_{x}-{\bar{\partial}}%
_{y}-(\partial _{x}-\partial _{y})h\right) \delta (x-y)\left[ \delta
(y-y_{1})\delta ^{\prime }(y-y_{2})-\delta (y-y_{2})\delta ^{\prime
}(y-y_{1})\right] ,  \notag \\
M_{22}^{hB}& =-2\alpha \left( 3B^{\prime }-2B\partial _{y}\right) \delta
(x-y)\left[ \delta (y-y_{1})\delta ^{\prime }(y-y_{2})-\delta
(y-y_{2})\delta ^{\prime }(y-y_{1})\right] ,  \label{an2}
\end{align}
depending on the fields $\phi $,$\;h$,$\;B$. The analytic choice of the
second-level structure functions (\ref{w3F}), (\ref{w3M}) is a particular
case of (\ref{an1}), (\ref{an2}), corresponding to $\alpha =0$.

Note that the extended analytic realization (\ref{an1}), (\ref{an2}) can be
obtained from the Ansatz (\ref{ansatz}) and the parameterization (\ref{param}%
), with $\alpha_{1}=1-\alpha$, $\beta_{1}=-1-\alpha$, $\alpha
_{2}=\beta_{2}=0$. These values of the parameters are related to a modified
Ansatz (\ref{ansatz2}), where $M_{22}^{\phi B}=0$, and a structure function $%
M_{22}^{hh}$ is included.

A remarkable property of the action (\ref{Acw3nBV}) is its dependence on the
ghost fields $c$, $l$. For any value of $\alpha $, they enter the action
only in the second order -- in contrast to the action (\ref{w3actsM}), which
depends on the ghost fields in the third order if $\beta \neq 0$.

\textbf{(d)} One can prove that the arbitrariness in analytic structure
functions described in \cite{VVanP} is unique in the sense that it preserves
the form of the action, being of second order in the ghost fields. Indeed,
to preserve a given set of gauge generators, the generating functional of
the anticanonical transformations must be at least of second order in the
antifields and ghost fields. It can be verified by straightforward
calculations that any simple form of such anticanonical transformations,
except (\ref{w3sct}), leads to an action depending on the ghost fields in
the third order. For example, let us consider an anticanonical
transformation with the generating functional 
\begin{equation*}
X(\phi ,\phi ^{\ast })=E(\phi ,\phi ^{\ast })-3\gamma \int d^{2}x\,B^{\ast
}c^{\ast }l^{\prime }l,
\end{equation*}
which is similar to (\ref{w3sct}). Then we obtain the action 
\begin{align}
& S(\gamma )=S_{0}+S_{1}+\int d^{2}x\,\Big[c^{\ast }\left( c^{\prime
}c+(1-\gamma \phi ^{\prime })\phi ^{\prime \,2}l^{\prime }l\right) +l^{\ast
}(l^{\prime }c+2c^{\prime }l)  \notag \\
& +3\gamma B^{\ast }\left( 2BB^{\ast \prime }-({\bar{\partial}}-h\partial
+h^{\prime })h^{\ast }\right) l^{\prime }l+\phi ^{\ast }\left( 2h^{\ast
}+3\gamma B^{\ast }\right) \phi ^{\prime }l^{\prime }l+3\gamma B^{\ast
}c^{\ast }c^{\prime }l^{\prime }l\Big],  \notag
\end{align}
being of the third order in the ghost fields and containing the structure
function $Q_{221}^{1B}$, which means that the Jacobi identity (\ref{GAnR})
for the gauge generators closes only on shell.

The above discussion shows that for $W_{3}$-gravity there exists a class of
``minimal'' actions, terminating at $S_{2}$ and related by anticanonical
transformations (cases a and c). Any other anticanonical transformation
produces higher ghost contributions, starting at $S_{3}$ (case d), and
leading also to non-analytic actions (case b). In what follows, we shall
consider only analytic realizations of the gauge algebra.

\section{$W_{3}$-gravity in the Sp(2)-covariant formalism}

Let us consider the model of $W_{3}$-gravity in the framework of the
Sp(2)-covariant extension \cite{blt} of the BV quantization scheme. To this
end, it is necessary to introduce the complete configuration space of fields 
$\phi ^{A}$, constructed from the classical fields, as well as from the
ghosts and auxiliary fields of the BV formalism, combined into completely
symmetric tensors under the group Sp(2). Thus, for irreducible theories the
fields $\phi ^{A}$ are given by 
\begin{equation}
\phi ^{A}=(A^{i},B^{\alpha },C^{\alpha a}),\;\;\;a=1,2,  \label{flds}
\end{equation}
where the Sp(2)-doublets $C^{\alpha a}$ stand for the ghost-antighost pairs $%
(C^{\alpha },\bar{C}^{\alpha })$, while the Sp(2)-scalars $B^{\alpha }$, $%
\varepsilon (B^{\alpha })=\varepsilon (\xi ^{\alpha })$, stand for the
Lagrange multipliers known as Nakanishi--Lautrup fields. The fields $\phi
^{A}$ are associated with the corresponding sets of antifields $\phi
_{Aa}^{\ast }$ and $\bar{\phi}_{A}$, 
\begin{equation*}
\varepsilon (\phi _{Aa}^{\ast })=\varepsilon (\phi ^{A})+1,\;\;\;\varepsilon
(\bar{\phi}_{A})=\varepsilon (\phi ^{A}).
\end{equation*}
Thus, the antifields corresponding to (\ref{flds}) are given by 
\begin{equation*}
\phi _{Aa}^{\ast }=(A_{ia}^{\ast },B_{\alpha a}^{\ast },C_{\alpha ab}^{\ast
}),\;\;\;\bar{\phi}_{A}=(\bar{A}_{i},\bar{B}_{\alpha },\bar{C}_{\alpha a}).
\end{equation*}
The fields $\phi ^{A}$ and antifields $\phi _{Aa}^{\ast }$, $\bar{\phi}_{A}$
are ascribed the so-called new ghost number, denoted by ``ngh'' and subject
to the following conditions: 
\begin{equation*}
\mathrm{ngh}(\phi _{Aa}^{\ast })=-1-\mathrm{ngh}(\phi ^{A}),\;\;\;\mathrm{ngh%
}(\bar{\phi}_{A})=-2-\mathrm{ngh}(\phi ^{A}),
\end{equation*}
where the new ghost number of the fields in the case (\ref{flds}) is given
by the rule 
\begin{equation*}
\mathrm{ngh}(A^{i})=0,\;\;\;\mathrm{ngh}(C^{\alpha a})=1,\;\;\;\mathrm{ngh}%
(B^{\alpha })=2.
\end{equation*}

The basic object of the Sp(2)-covariant formalism \cite{blt} is a bosonic
functional $S=S(\phi,\phi^{\ast},\bar{\phi})$ subject to the classical
master equation

\begin{equation}
\frac{\delta S}{\delta\phi^{A}}\frac{\delta S}{\delta\phi_{Aa}^{\ast}}%
+\varepsilon^{ab}\phi_{Ab}^{\ast}\frac{\delta S}{\delta\bar{\phi}_{A}}=0,
\label{sp(2)_eq}
\end{equation}
with the boundary condition 
\begin{equation*}
\left. S\right| _{\bar{\phi}=\phi^{\ast}=0}=S_{0}(A),
\end{equation*}
where $\varepsilon^{ab}$ is a constant antisymmetric second-rank tensor,
with $\varepsilon^{ac}\varepsilon_{cb}=\delta_{b}^{a}$ and $%
\varepsilon^{12}=1$. The existence of solutions of CME in the
Sp(2)-covariant formalism has been proved \cite{blt} for both irreducible
and reducible gauge theories of general kind. These solutions are sought as
series in ghost and auxiliary fields $(C,B)$, under the requirement of the
new ghost number conservation: 
\begin{equation}
S=S_{0}+\sum_{n=1}S_{n},\;\;\;\varepsilon(S_{n})=\mathrm{ngh}%
(S_{n})=0,\;\;\;S_{n}\thicksim(C)^{n-m}(B)^{m},\;\;\;0\leq m\leq n.
\label{series}
\end{equation}

For irreducible theories of general kind, an approximated solution of CME
was found \cite{lmr} up to the third order in the powers of ghosts and
auxiliary fields $(C^{\alpha a},B^{\alpha })$. The approximation found in 
\cite{lmr} is completely determined by the structure functions up to the
third level. %
%
%
%
%
%

In the bosonic case $\varepsilon (A^{i})=\varepsilon (\xi ^{\alpha })=0$,
the solution found in \cite{lmr} has the form 
\begin{align}
& S(\phi ,\phi ^{\ast },\bar{\phi})=S_{0}(A)+A_{ia}^{\ast }R_{\alpha
}^{i}C^{\alpha a}+\bar{A}_{i}R_{\alpha }^{i}B^{\alpha }-\varepsilon
^{ab}C_{\alpha ab}^{\ast }B^{\alpha }-\frac{1}{2}C_{\gamma ab}^{\ast
}F_{\alpha \beta }^{\gamma }C^{\beta b}C^{\alpha a}  \notag \\
& +\frac{1}{4}A_{ia}^{\ast }A_{jb}^{\ast }M_{\alpha \beta }^{ij}C^{\beta
b}C^{\alpha a}+\frac{1}{2}\bar{A}_{i}R_{\alpha ,j}^{i}R_{\beta }^{j}C^{\beta
b}C^{\alpha a}\varepsilon _{ab}+\frac{1}{2}(2\bar{C}_{\gamma a}-B_{\gamma
a}^{\ast })F_{\alpha \beta }^{\gamma }B^{\beta }C^{\alpha a}+\frac{1}{2}%
A_{ia}^{\ast }\bar{A}_{j}M_{\alpha \beta }^{ij}B^{\beta }C^{\alpha a}  \notag
\\
& +\frac{1}{12}\left( 2\bar{C}_{\delta b}-B_{\delta b}^{\ast }\right) \left(
F_{\alpha \sigma }^{\delta }F_{\beta \gamma }^{\sigma }+2F_{\alpha \beta
,i}^{\delta }R_{\gamma }^{i}\right) C^{\gamma c}C^{\beta b}C^{\alpha
a}\varepsilon _{ac}-\frac{1}{2}\bar{A}_{i}\bar{A}_{j}R_{\alpha
,k}^{i}M_{\beta \gamma }^{jk}B^{\gamma }C^{\beta b}C^{\alpha a}\varepsilon
_{ab}  \notag \\
& -\frac{1}{12}A_{ib}^{\ast }\bar{A}_{j}\left( 2R_{\alpha ,k}^{i}M_{\beta
\gamma }^{jk}+4R_{\alpha ,k}^{j}M_{\beta \gamma }^{ik}-M_{\alpha \delta
}^{ij}F_{\beta \gamma }^{\delta }-2M_{\alpha \beta ,k}^{ij}R_{\gamma
}^{k}\right) C^{\gamma c}C^{\beta b}C^{\alpha a}\varepsilon _{ac}+...,
\label{s_sp(2)}
\end{align}
where we have assumed the absence of higher structure functions, as in the
case of $W_{3}$-gravity with the one-parametric family of analytic
realizations of $F_{\alpha \beta }^{\gamma }$ and $M_{\alpha \beta }^{ij}$,
given by (\ref{an1}), (\ref{an2}).

Let us calculate the approximated expression (\ref{s_sp(2)}) in the case of $%
W_{3}$-gravity with the simplest analytic choice of the second-level
structure functions (\ref{w3F}), (\ref{w3M}), corresponding to the zero
value of the arbitrary parameter in (\ref{an1}), (\ref{an2}). Denoting the
ghosts and auxiliary fields in $W_{3}$-gravity by $C^{\alpha
a}=(c^{a},l^{a}) $ and $B^{\alpha }=(u,v)$, we shall write down the
contributions $S_{1}$, $S_{2}$, $S_{3}$ corresponding to (\ref{s_sp(2)}).

In the first order: 
\begin{align}
S_{1}& =\int d^{2}x\Big[\phi _{a}^{\ast }\left( \phi ^{\prime }c^{a}+\phi
^{\prime \,2}l^{a}\right) +h_{a}^{\ast }\left( \bar{\nabla}_{1}c^{a}-\phi
^{\prime \,2}\nabla _{1}(Bl^{a})\right) +B_{a}^{\ast }\left( \bar{\nabla}%
_{2}l^{a}-2\nabla _{\frac{1}{2}}(Bc^{a})\right)  \notag \\
& +{\bar{\phi}}(\phi ^{\prime }u+\phi ^{\prime \,2}v)+{\bar{h}}\left( \bar{%
\nabla}_{1}u-\phi ^{\prime \,2}\nabla _{1}(Bv)\right) +{\bar{B}}\left( \bar{%
\nabla}_{2}v-2\nabla _{\frac{1}{2}}(Bu)\right) -\varepsilon
^{ab}c_{ab}^{\ast }u-\varepsilon ^{ab}l_{ab}^{\ast }v\Big].  \label{s1}
\end{align}
In the second order:

\begin{align}
S_{2}& =\int d^{2}x\bigg\{-\frac{1}{2}c_{ab}^{\ast }\left[ \nabla
_{1}(c^{b}c^{a})+\phi ^{\prime 2}\nabla _{1}(l^{b}l^{a})\right] -\frac{1}{2}%
l_{ab}^{\ast }\left[ \nabla _{2}(c^{b}l^{a})+2\nabla _{\frac{1}{2}%
}(l^{b}c^{a})\right]  \notag \\
& +\left( \frac{1}{2}u_{a}^{\ast }-\bar{c}_{a}\right) \left[ \nabla
_{1}(c^{a}u)+\phi ^{\prime 2}\nabla _{1}(l^{a}v)\right] +\left( \frac{1}{2}%
v_{a}^{\ast }-\bar{l}_{a}\right) \left[ \nabla _{2}(c^{a}v)+2\nabla _{\frac{1%
}{2}}(l^{a}u)\right]  \notag \\
& -\phi _{a}^{\ast }h_{b}^{\ast }\phi ^{\prime }\nabla _{1}(l^{b}l^{a})+(%
\bar{h}\phi _{a}^{\ast }-{\bar{\phi}}h_{a}^{\ast })\phi ^{\prime }\nabla
_{1}(l^{a}v)+\frac{1}{2}\varepsilon _{ab}{\bar{\phi}}(c^{b}+2\phi ^{\prime
}l^{b})\left( \phi ^{\prime }c^{a}+\phi ^{\prime \,2}l^{a}\right) ^{\prime }
\notag \\
& +\frac{1}{2}\varepsilon _{ab}\bar{h}\bigg[2\phi ^{\prime }(\phi ^{\prime
}c^{a}+\phi ^{\prime \,2}l^{a})^{\prime }\nabla _{1}(Bl^{b})-\phi ^{\prime
\,2}\nabla _{1}\left[ \left( \bar{\nabla}_{2}l^{a}-2\nabla _{\frac{1}{2}%
}(Bc^{a})\right) l^{b}\right]  \notag \\
& +\nabla _{1}\left[ \left( \bar{\nabla}_{1}c^{a}-\phi ^{\prime \,2}\nabla
_{1}(Bl^{a})\right) c^{b}\right] \bigg]-\frac{1}{2}\varepsilon _{ab}{\bar{B}}%
\bigg[\nabla _{2}\left[ \left( \bar{\nabla}_{2}l^{a}-2\nabla _{\frac{1}{2}%
}(Bc^{a})\right) c^{b}\right]  \notag \\
& -\nabla _{2}\left[ \left( \bar{\nabla}_{1}c^{a}-\phi ^{\prime \,2}\nabla
_{1}(Bl^{a})\right) l^{b}\right] \bigg]\bigg\}.  \label{s2}
\end{align}
In the third order:

\begin{align}
S_{3}& =\int d^{2}x\bigg\{\frac{1}{6}\varepsilon _{cd}\left( \frac{1}{2}%
u_{a}^{\ast }-\bar{c}_{a}\right) \bigg[\nabla _{1}\left[ \left( \nabla
_{1}(c^{c}c^{a})+\phi ^{\prime \,2}\nabla _{1}(l^{c}l^{a})\right) c^{d}%
\right]  \notag \\
& +\phi ^{\prime \,2}\nabla _{1}\left[ \left( \nabla
_{2}(c^{c}l^{a})+2\nabla _{\frac{1}{2}}(l^{c}c^{a})\right) l^{d}\right]
+4\phi ^{\prime }(\phi ^{\prime }c^{c}+\phi ^{\prime \,2}l^{c})^{\prime
}\nabla _{1}(l^{a}l^{d})\bigg]  \notag \\
& +\frac{1}{6}\varepsilon _{cd}\left( \frac{1}{2}v_{a}^{\ast }-\bar{l}%
_{a}\right) \Big\{\nabla _{2}\left[ \left( \nabla _{1}(c^{c}c^{a})+\phi
^{\prime \,2}\nabla _{1}(l^{c}l^{a})\right) l^{d}\right]  \notag \\
& -\nabla _{2}\left[ \left( \nabla _{2}(c^{c}l^{a})+2\nabla _{\frac{1}{2}%
}(l^{c}c^{a})\right) c^{d}\right] \Big\}-\bar{\phi}\varepsilon _{cd}\left[ 
\bar{h}\phi ^{\prime }\nabla _{1}(l^{c}v)\right] ^{\prime }(c^{d}+2\phi
^{\prime }l^{d})  \notag \\
& -\frac{1}{2}\varepsilon _{cd}\bar{h}\bigg[\nabla _{1}\left[ \left( \bar{%
\phi}\phi ^{\prime }\nabla _{1}(l^{c}v)\right) c^{d}\right] -2\bar{h}\phi
^{\prime }[\nabla _{1}(l^{c}v)]^{\prime }\nabla _{1}(Bl^{d})\bigg]  \notag \\
& -\varepsilon _{cd}\bar{B}\nabla _{2}\left[ \left( \bar{\phi}\phi ^{\prime
}\nabla _{1}(l^{c}v)\right) l^{d}\right] +\frac{1}{3}\varepsilon _{cd}\phi
_{a}^{\ast }\left[ \bar{h}\phi ^{\prime }\nabla _{1}(l^{c}l^{a})\right]
^{\prime }(c^{d}+2\phi ^{\prime }l^{d})  \notag \\
& +\frac{1}{3}\varepsilon _{cd}h_{a}^{\ast }\left[ \nabla _{1}\left[ \left( 
\bar{\phi}\phi ^{\prime }\nabla _{1}(l^{c}l^{a})\right) c^{d}\right] -2\phi
^{\prime }\left[ \bar{h}\phi ^{\prime }\nabla _{1}(l^{c}l^{a})\right]
^{\prime }\nabla _{1}(Bl^{d})\right]  \notag \\
& +\frac{1}{3}\varepsilon _{cd}B_{a}^{\ast }\nabla _{2}\left[ \left( \bar{%
\phi}\phi ^{\prime }\nabla _{1}(l^{c}l^{a})\right) l^{d}\right] +\frac{2}{3}%
\varepsilon _{cd}\bar{\phi}\left[ h_{a}^{\ast }\phi ^{\prime }\nabla
_{1}\left( l^{c}l^{a}\right) \right] ^{\prime }(c^{d}+2\phi ^{\prime }l^{d})
\notag \\
& +\frac{2}{3}\varepsilon _{cd}\bar{h}\left[ \nabla _{1}\left[ \left( \phi
_{a}^{\ast }\phi ^{\prime }\nabla _{1}(l^{c}l^{a})\right) c^{d}\right]
-2\phi ^{\prime }\left[ h_{a}^{\ast }\phi ^{\prime }\nabla _{1}(l^{c}l^{a})%
\right] ^{\prime }\nabla _{1}(Bl^{d})\right]  \notag \\
& +\frac{2}{3}\varepsilon _{cd}\bar{B}\nabla _{2}\left[ \left( \phi
_{a}^{\ast }\phi ^{\prime }\nabla _{1}(l^{c}l^{a})\right) l^{d}\right] -%
\frac{1}{6}\varepsilon _{cd}(h_{a}^{\ast }\bar{\phi}-\phi _{a}^{\ast }\bar{h}%
)\bigg[\phi ^{\prime }\nabla _{1}\Big[\Big(\nabla _{2}(c^{c}l^{a})  \notag \\
& +2\nabla _{\frac{1}{2}}(l^{c}c^{a})\Big)l^{d}\Big]+2(\phi ^{\prime
}c^{c}+\phi ^{\prime \,2}l^{c})^{\prime }\nabla _{1}(l^{a}l^{d})\bigg]\bigg\}%
.  \label{s3}
\end{align}
In (\ref{s1}), (\ref{s2}), (\ref{s3}), we have used the following notation: 
\begin{equation}
\nabla _{j}(XY)\equiv XY^{\prime }-jX^{\prime }Y,\;\;\;\bar{\nabla}%
_{j}X\equiv \dot{X}-\nabla _{j}(hX),  \label{alavan}
\end{equation}
where $j$ is a real number.

The details of calculations are presented in Appendix A, using the example
of $S_{3}$. Contrary to the expectations motivated by the results of the BV
formalism, the approximation given by (\ref{s1}), (\ref{s2}), (\ref{s3})
does not provide a closed solution of CME, which is shown in Appendix B.

Therefore, a closed solution must contain higher contributions $S_{n}$. To
tackle this problem, one has to deal with the task of finding further
approximations for the Sp(2)-covariant extended action of general
irreducible theories with open algebras. This problem remains open. However,
since the complexity of the approximations will gradually increase, it is
not evident that a finite number of contributions will be sufficient to
provide a solution in the given example of $W_{3}$-gravity. Concerning the
possibility of a closed solution with a finite number of contributions,
there is evidence that such a solution may require contributions up to $%
S_{6} $ (see Appendix B). In this case, the task of solving CME seems to be
extremely difficult. One could have hoped that the problem might be attacked
with the help of some transformations analogous to anticanonical ones. The
role of such transformations in the Sp(2)-covariant scheme is played by
operator transformations of the quantum action \cite{blt}. However, it
should be noted that we have already started from the simplest realization
of the gauge algebra, which in the BV formalism immediately provides a
closed solution.

As mentioned previously, the analyzed example of a gauge theory with an open
algebra is relatively simple, and therefore in the case of more complicated
theories with open algebras it is natural to expect more technical
difficulties. Nevertheless, the possibility of sufficient conditions that
ensure the existence of a closed solution of the Sp(2)-covariant CME for
theories with open algebras is an interesting problem that deserves
investigation.

Concluding, note that in the limit $B=$ $l^{a}=v=0$ the functional $%
S_{0}+S_{1}+S_{2}+S_{3}$ coincides with the closed solution \cite{glm} of
CME for the model of $W_{2}$-gravity \cite{pol}, whose classical action is
given by (\ref{w3act}) in the limit $B=0$, and whose gauge transformations
are given by (\ref{g_tr}) in the limit $\lambda =0$.

{\Large \textbf{Acknowledgments }}B.G. gratefully acknowledges the support
of the German--Brazil scientific exchange programmes DAAD and FAPESP, as
well as the hospitality of the Institute of Physics, University of S\~{a}o
Paulo. D.M.G., P.M.L. and P.Yu.M. acknowledge the hospitality of NTZ at the
Center of Advanced Study, Leipzig University. D.M.G. is grateful to the
foundations FAPESP, CNPq and DAAD for support. The work of P.M.L. was
supported by INTAS, grant 99-0590, as well as by the project of Deutsche
Forschunsgemeinschaft (DFG), 436 RUS 113/669, and Russian Foundation for
Basic Research (RFBR), 02-02-04002. The work of P.Yu.M. was supported by the
Funda\c{c}\~{a}o de Amparo \`{a} Pesquisa do Estado de S\~{a}o Paulo
(FAPESP), grant 02/00423-4.

{\Large \textbf{Appendix A}}

\setcounter{equation}{0} \renewcommand{\theequation}{A.\arabic{equation}} In
this Appendix, we shall calculate the contribution $S_{3}$, (\ref{s3}),
using a technique of reducing the corresponding expression to gauge algebra
operations. The same technique can be applied to the calculation of the
previous contributions $S_{1}$, $S_{2}$, (\ref{s1}), (\ref{s2}).

Let us consider the part of (\ref{s_sp(2)}), corresponding to $S_{3}$, 
\begin{align}
S_{3}& =\frac{1}{12}(2\bar{C}_{\delta b}-B_{\delta b}^{\ast })(F_{\alpha
\sigma }^{\delta }F_{\beta \gamma }^{\sigma }+2F_{\alpha \beta ,i}^{\delta
}R_{\gamma }^{i})C^{\gamma c}C^{\beta b}C^{\alpha a}\varepsilon _{ac}-\frac{1%
}{2}\bar{A}_{i}\bar{A}_{j}R_{\alpha ,k}^{i}M_{\beta \gamma }^{jk}B^{\gamma
}C^{\beta b}C^{\alpha a}\varepsilon _{ab}  \notag \\
& -\frac{1}{12}A_{ib}^{\ast }\bar{A}_{j}(2R_{\alpha ,k}^{i}M_{\beta \gamma
}^{jk}+4R_{\alpha ,k}^{j}M_{\beta \gamma }^{ik}-M_{\alpha \delta
}^{ij}F_{\beta \gamma }^{\delta }-2M_{\alpha \beta ,k}^{ij}R_{\gamma
}^{k})C^{\gamma c}C^{\beta b}C^{\alpha a}\varepsilon _{ac}\,.
\label{s3_sp(2)}
\end{align}

To simplify the consideration of this functional, let us introduce a set of
three constant Grassmann doublets $\mu _{a(k)}$, $k=1,2,3$, and a set of
four bosonic parameters $\xi _{(n)}^{\alpha }$, $n=0,1,2,3$, by the rule 
\begin{equation}
\xi _{(n)}^{\alpha }=(\xi _{(0)}^{\alpha },\xi _{(k)}^{\alpha }),\;\;\;\xi
_{(0)}^{\alpha }\equiv B^{\alpha },\;\;\;\xi _{(k)}^{\alpha }\equiv
C^{\alpha a}\mu _{a(k)}.  \label{g_param}
\end{equation}
With this notation, we have\footnote{%
The derivatives are applied in the following order:\ $\frac{\partial _{r}}{%
\partial \mu _{b(2)}}\frac{\partial _{r}}{\partial \mu _{a(1)}}F=\frac{%
\partial _{r}}{\partial \mu _{b(2)}}\left( \frac{\partial _{r}}{\partial \mu
_{a(1)}}F\right) $.} 
\begin{equation}
S_{3}=\varepsilon _{ab}\frac{\partial _{r}}{\partial \mu _{b(2)}}\frac{%
\partial _{r}}{\partial \mu _{a(1)}}F+\varepsilon _{ac}\frac{\partial _{r}}{%
\partial \mu _{c(3)}}\frac{\partial _{r}}{\partial \mu _{b(2)}}\frac{%
\partial _{r}}{\partial \mu _{a(1)}}F_{b}\,,  \label{s3new}
\end{equation}
where $F$ and $F_{a}$ are given by 
\begin{equation*}
F=\frac{1}{2}\bar{A}_{i}\bar{A}_{j}M_{\alpha \beta }^{jk}\xi _{(0)}^{\beta
}\xi _{(1)}^{\alpha }(R_{\gamma }^{i}\xi _{(2)}^{\gamma })_{,k}
\end{equation*}
and 
\begin{align*}
F_{a}& =-\frac{1}{12}(2\bar{C}_{\delta a}-B_{\delta a}^{\ast })\left[
F_{\gamma \sigma }^{\delta }F_{\beta \alpha }^{\sigma }\xi _{(1)}^{\alpha
}\xi _{(2)}^{\beta }\xi _{(3)}^{\gamma }+2R_{\gamma }^{i}\xi _{(1)}^{\gamma
}(F_{\alpha \beta }^{\delta }\xi _{(2)}^{\beta }\xi _{(3)}^{\alpha })_{,i}%
\right] \\
& +\frac{1}{6}\left( A_{ia}^{\ast }\bar{A}_{j}M_{\alpha \beta }^{jk}\xi
_{(1)}^{\beta }\xi _{(2)}^{\alpha }+2\bar{A}_{i}A_{ja}^{\ast }M_{\alpha
\beta }^{jk}\xi _{(1)}^{\beta }\xi _{(2)}^{\alpha }\right) (R_{\gamma
}^{i}\xi _{(3)}^{\gamma })_{,k} \\
& -\frac{1}{12}A_{ia}^{\ast }\bar{A}_{j}\left[ M_{\gamma \delta
}^{ij}F_{\alpha \beta }^{\delta }\xi _{(1)}^{\alpha }\xi _{(2)}^{\beta }\xi
_{(3)}^{\gamma }+2R_{\gamma }^{k}\xi _{(1)}^{\gamma }(M_{\alpha \beta
}^{ij}\xi _{(2)}^{\beta }\xi _{(3)}^{\alpha })_{,k}\right] .
\end{align*}
The above expressions can be rewritten in the compact form 
\begin{equation}
F=\frac{1}{2}\bar{A}_{i}\left. \delta (\delta _{(2)}A^{i})\right| _{\delta
A\rightarrow \bar{\delta}A_{(0,1)}}  \label{a1}
\end{equation}
and 
\begin{align}
F_{a}& =-\frac{1}{12}(2\bar{C}_{\gamma a}-B_{\gamma a}^{\ast })(\tilde{\xi}%
_{((1,2),3)}^{\gamma }+2\delta _{(1)}\tilde{\xi}_{(2,3)}^{\gamma })+\frac{1}{%
6}A_{ia}^{\ast }\left. \delta (\delta _{(3)}A^{i})\right| _{\delta
A\rightarrow \bar{\delta}A_{(1,2)}}  \notag \\
& +\frac{1}{3}\bar{A}_{i}\left. \delta (\delta _{(3)}A^{i})\right| _{\delta
A\rightarrow \delta _{a}^{\ast }A_{(1,2)}}-\frac{1}{12}A_{ia}^{\ast }\bar{A}%
_{j}(\bar{\xi}_{((1,2),3)}^{ij}+2\delta _{(1)}\bar{\xi}_{(2,3)}^{ij}),
\label{a2}
\end{align}
where $\delta _{(n)}$ stand for gauge variations with parameters $\xi
_{(n)}^{\alpha }$; the quantities $\tilde{\xi}_{(m,n)}^{\gamma }$, $\tilde{%
\xi}_{((m,n),l)}^{\gamma }$ are defined by 
\begin{equation}
\tilde{\xi}_{(m,n)}^{\gamma }=F_{\alpha \beta }^{\gamma }\xi _{(m)}^{\beta
}\xi _{(n)}^{\alpha },\;\;\;\tilde{\xi}_{((m,n),l)}^{\gamma }=F_{\alpha
\beta }^{\gamma }\tilde{\xi}_{(m,n)}^{\beta }\xi _{(l)}^{\alpha };
\label{param4}
\end{equation}
the variations $\delta $ are understood as usual variations of the
quantities $\delta _{(n)}A^{i}$ with respect to $A^{i}$, where $\delta A^{i}$
in the resultant expression $\delta (\delta _{(n)}A^{i})$ are replaced by $%
\bar{\delta}A_{(m,n)}^{i}$, $\delta _{a}^{\ast }A_{(m,n)}^{i}$, having the
form 
\begin{equation*}
\bar{\delta}A_{(m,n)}^{i}=\bar{A}_{j}\bar{\xi}_{(m,n)}^{ji},\;\;\;\delta
_{a}^{\ast }A_{(m,n)}^{i}=A_{ja}^{\ast }\bar{\xi}_{(m,n)}^{ji},
\end{equation*}
in accordance with the definition of $\bar{\xi}_{(m,n)}^{ij}$ and $\bar{\xi}%
_{((m,n),l)}^{ij}$: 
\begin{equation}
\bar{\xi}_{(m,n)}^{ij}\equiv M_{\alpha \beta }^{ij}\xi _{(m)}^{\beta }\xi
_{(n)}^{\alpha },\;\;\;\bar{\xi}_{((m,n),l)}^{ij}\equiv M_{\alpha \beta
}^{ij}\tilde{\xi}_{(m,n)}^{\beta }\xi _{(l)}^{\alpha }.  \label{param5}
\end{equation}

To calculate the quantities $F$ and $F_{a}$ in the model of $W_{3}$-gravity,
we remind that $A^{i}=(\phi ,h,B)$, $\xi ^{\alpha }=(\epsilon ,\lambda )$,
with the gauge transformations $\delta A^{i}=R_{\alpha }^{i}\xi ^{\alpha }$
given by (\ref{g_tr}). The (non-vanishing) quantities $\tilde{\xi}%
_{(m,n)}^{\alpha }$, $\bar{\xi}_{(m,n)}^{ij}$, corresponding to subsequent
gauge transformations with the parameters $\xi _{(m)}^{\alpha }$ and $\xi
_{(n)}^{\alpha }$, have the form 
\begin{equation}
\tilde{\xi}_{(m,n)}^{\alpha }=F_{\beta \gamma }^{\alpha }\xi _{(m)}^{\gamma
}\xi _{(n)}^{\beta }=(\tilde{\epsilon}_{(m,n)},\tilde{\lambda}%
_{(m,n)}),\;\;\;\bar{\xi}_{(m,n)}^{ij}=M_{\beta \gamma }^{ij}\xi
_{(m)}^{\gamma }\xi _{(n)}^{\beta }=(\bar{\xi}_{(m,n)}^{\phi h},\bar{\xi}%
_{(m,n)}^{h\phi }),  \label{param7}
\end{equation}
with 
\begin{align}
\tilde{\epsilon}_{(m,n)}& =\epsilon _{(m)}\epsilon _{(n)}^{\prime }-\epsilon
_{(m)}^{\prime }\epsilon _{(n)}+\phi ^{\prime \,2}(\lambda _{(m)}\lambda
_{(n)}^{\prime }-\lambda _{(m)}^{\prime }\lambda _{(n)}),  \notag \\
\tilde{\lambda}_{(m,n)}& =\epsilon _{(m)}\lambda _{(n)}^{\prime }-2\epsilon
_{(m)}^{\prime }\lambda _{(n)}-\lambda _{(m)}^{\prime }\epsilon
_{(n)}+2\lambda _{(m)}\epsilon _{(n)}^{\prime },  \notag \\
\bar{\xi}_{(m,n)}^{\phi h}& =-\bar{\xi}_{(m,n)}^{h\phi }=-2\phi ^{\prime
}(\lambda _{(m)}\lambda _{(n)}^{\prime }-\lambda _{(m)}^{\prime }\lambda
_{(n)}),  \label{param8}
\end{align}
which follows from the parameterization (\ref{param2}), (\ref{param3}), (\ref
{param}) of the second-level structure functions in the case $\alpha
_{1}=1,\;\alpha _{2}=0$, $\beta _{1}=-1$, $\beta _{2}=0$, corresponding to
the choice of $F_{\alpha \beta }^{\gamma }$, $M_{\alpha \beta }^{ij}$ in the
analytic form (\ref{w3F}), (\ref{w3M}).

Using the above identifications, we are now able to calculate all the
structures which enter the quantities $F$ and $F_{a}$: 
\begin{gather}
\tilde{\xi}_{((m,n),l)}^{\alpha },\;\;\;\bar{\xi}_{((m,n),l)}^{ij},\;\;\;%
\delta _{(l)}\tilde{\xi}_{(m,n)}^{\alpha },\;\;\;\delta _{(l)}\bar{\xi}%
_{(m,n)}^{ij},  \notag \\
\left. \delta (\delta _{(l)}A_{ia}^{\ast }A^{i})\right| _{\delta
A\rightarrow \bar{\delta}A_{(m,n)}},\;\;\;\left. \delta (\delta _{(l)}\bar{A}%
_{i}A^{i})\right| _{\delta A\rightarrow \delta _{a}^{\ast
}A_{(m,n)}},\;\;\;\left. \delta (\delta _{(l)}\bar{A}_{i}A^{i})\right|
_{\delta A\rightarrow \bar{\delta}A_{(m,n)}}.  \label{struct}
\end{gather}
The structure $\tilde{\xi}_{((m,n),l)}^{\alpha }$ is given by 
\begin{equation}
\tilde{\xi}_{((m,n),l)}^{\alpha }=(\tilde{\epsilon}_{((m,n),l)},\tilde{%
\lambda}_{((m,n),l)}),  \label{param9}
\end{equation}
where 
\begin{align*}
\tilde{\epsilon}_{((m,n),l)}& =\tilde{\epsilon}_{(m,n)}\epsilon
_{(l)}^{\prime }-\tilde{\epsilon}_{(m,n)}^{\prime }\epsilon _{(l)}+\phi
^{\prime \,2}(\tilde{\lambda}_{(m,n)}\lambda _{(l)}^{\prime }-\tilde{\lambda}%
_{(m,n)}^{\prime }\lambda _{(l)})= \\
& =[\epsilon _{(m)}\epsilon _{(n)}^{\prime }-\epsilon _{(m)}^{\prime
}\epsilon _{(n)}+\phi ^{\prime \,2}(\lambda _{(m)}\lambda _{(n)}^{\prime
}-\lambda _{(m)}^{\prime }\lambda _{(n)})]\epsilon _{(l)}^{\prime }- \\
& -[\epsilon _{(m)}\epsilon _{(n)}^{\prime }-\epsilon _{(m)}^{\prime
}\epsilon _{(n)}+\phi ^{\prime \,2}(\lambda _{(m)}\lambda _{(n)}^{\prime
}-\lambda _{(m)}^{\prime }\lambda _{(n)})]^{\prime }\epsilon _{(l)}^{\prime
}+ \\
& +\phi ^{\prime \,2}(\epsilon _{(m)}\lambda _{(n)}^{\prime }-2\epsilon
_{(m)}^{\prime }\lambda _{(n)}-\lambda _{(m)}^{\prime }\epsilon
_{(n)}+2\lambda _{(m)}\epsilon _{(n)}^{\prime })\lambda _{(l)}^{\prime }- \\
& -\phi ^{\prime \,2}(\epsilon _{(m)}\lambda _{(n)}^{\prime }-2\epsilon
_{(m)}^{\prime }\lambda _{(n)}-\lambda _{(m)}^{\prime }\epsilon
_{(n)}+2\lambda _{(m)}\epsilon _{(n)}^{\prime })^{\prime }\lambda _{(l)}, \\
\tilde{\lambda}_{((m,n),l)}& =\tilde{\epsilon}_{(m,n)}\partial \lambda
_{(l)}-2\tilde{\epsilon}_{(m,n)}^{\prime }\lambda _{(l)}-\tilde{\lambda}%
_{(m,n)}^{\prime }\epsilon _{(l)}+2\tilde{\lambda}_{(m,n)}\epsilon
_{(l)}^{\prime }= \\
& =[\epsilon _{(m)}\epsilon _{(n)}^{\prime }-\epsilon _{(m)}^{\prime
}\epsilon _{(n)}+\phi ^{\prime \,2}(\lambda _{(m)}\lambda _{(n)}^{\prime
}-\lambda _{(m)}^{\prime }\lambda _{(n)})]\lambda _{(l)}^{\prime }- \\
& -2[\epsilon _{(m)}\epsilon _{(n)}^{\prime }-\epsilon _{(m)}^{\prime
}\epsilon _{(n)}+\phi ^{\prime \,2}(\lambda _{(m)}\lambda _{(n)}^{\prime
}-\lambda _{(m)}^{\prime }\lambda _{(n)})]^{\prime }\lambda _{(l)}- \\
& -(\epsilon _{(m)}\lambda _{(n)}^{\prime }-2\epsilon _{(m)}^{\prime
}\lambda _{(n)}-\lambda _{(m)}^{\prime }\epsilon _{(n)}+2\lambda
_{(m)}\epsilon _{(n)}^{\prime })^{\prime }\epsilon _{(l)}+ \\
& +2(\epsilon _{(m)}\lambda _{(n)}^{\prime }-2\epsilon _{(m)}^{\prime
}\lambda _{(n)}-\lambda _{(m)}^{\prime }\epsilon _{(n)}+2\lambda
_{(m)}\epsilon _{(n)}^{\prime })\epsilon _{(l)}^{\prime }.
\end{align*}
To calculate $\delta _{(l)}\tilde{\xi}_{(m,n)}^{\alpha }$, we notice that 
\begin{equation*}
\delta _{(l)}\tilde{\xi}_{(m,n)}^{\alpha }=(\delta _{(l)}\tilde{\epsilon}%
_{(m,n)},\delta _{(l)}\tilde{\lambda}_{(m,n)}),
\end{equation*}
where 
\begin{equation*}
\delta _{(l)}\tilde{\epsilon}_{(m,n)}=2\phi ^{\prime }\left( \phi ^{\prime
}\epsilon _{(l)}+\phi ^{\prime \,2}\lambda _{(l)}\right) ^{\prime }(\lambda
_{(m)}\lambda _{(n)}^{\prime }-\lambda _{(m)}^{\prime }\lambda
_{(n)}),\;\;\;\delta _{(l)}\tilde{\lambda}_{(m,n)}=0.
\end{equation*}
Similarly, we determine the manifest form of $\bar{\xi}_{((m,n),l)}^{ij}$, 
\begin{align}
& \bar{\xi}_{((m,n),l)}^{ij}=(\bar{\xi}_{((m,n),l)}^{\phi h},\bar{\xi}%
_{((m,n),l)}^{h\phi }),  \notag \\
& \bar{\xi}_{((m,n),l)}^{\phi h}=-\bar{\xi}_{((m,n),l)}^{h\phi }=-2\phi
^{\prime }(\epsilon _{(m)}\lambda _{(n)}^{\prime }-2\epsilon _{(m)}^{\prime
}\lambda _{(n)}-\lambda _{(m)}^{\prime }\epsilon _{(n)}+2\lambda
_{(m)}\epsilon _{(n)}^{\prime })\lambda _{(l)}^{\prime }  \notag \\
& +2\phi ^{\prime }(\epsilon _{(m)}\lambda _{(n)}^{\prime }-2\epsilon
_{(m)}^{\prime }\lambda _{(n)}-\lambda _{(m)}^{\prime }\epsilon
_{(n)}+2\lambda _{(m)}\epsilon _{(n)}^{\prime })^{\prime }\lambda _{(l)},
\label{xi3}
\end{align}
and the structure $\delta _{(l)}\bar{\xi}_{(m,n)}^{ij}$, 
\begin{equation*}
\delta _{(l)}\bar{\xi}_{(m,n)}^{ij}=(\delta _{(l)}\bar{\xi}_{(m,n)}^{\phi
h},\delta _{(l)}\bar{\xi}_{(m,n)}^{h\phi }),
\end{equation*}
which implies

\begin{equation}
\delta _{(l)}\bar{\xi}_{(m,n)}^{\phi h}=-2(\phi ^{\prime }\epsilon
_{(l)}+\phi ^{\prime \,2}\lambda _{(l)})^{\prime }(\lambda _{(m)}\lambda
_{(n)}^{\prime }-\lambda _{(m)}^{\prime }\lambda _{(n)}).  \label{varxi}
\end{equation}
To calculate the quantities 
\begin{equation*}
\left. \delta (\delta _{(l)}A_{ia}^{\ast }A^{i})\right| _{\delta
A\rightarrow \bar{\delta}A_{(m,n)}},\;\;\;\left. \delta (\delta _{(l)}\bar{A}%
_{i}A^{i})\right| _{\delta A\rightarrow \delta _{a}^{\ast
}A_{(m,n)}},\;\;\;\left. \delta (\delta _{(l)}\bar{A}_{i}A^{i})\right|
_{\delta A\rightarrow \bar{\delta}A_{(m,n)}}\,,
\end{equation*}
we notice that the field variations are given by 
\begin{align*}
\bar{\delta}A_{(m,n)}^{i}& =\bar{A}_{j}\bar{\xi}_{(m,n)}^{ji}=\bar{\phi}\bar{%
\xi}_{(m,n)}^{\phi i}+\bar{h}\bar{\xi}_{(m,n)}^{hi}=(\bar{h}\bar{\xi}%
_{(m,n)}^{h\phi },\bar{\phi}\bar{\xi}_{(m,n)}^{\phi h},0) \\
\delta _{a}^{\ast }A_{(m,n)}^{i}& =A_{ja}^{\ast }\bar{\xi}_{(m,n)}^{ji}=\phi
_{a}^{\ast }\bar{\xi}_{(m,n)}^{\phi i}+h_{a}^{\ast }\bar{\xi}%
_{(m,n)}^{hi}=(h_{a}^{\ast }\bar{\xi}_{(m,n)}^{h\phi },\phi _{a}^{\ast }\bar{%
\xi}_{(m,n)}^{\phi h},0),
\end{align*}
and therefore 
\begin{align}
\bar{\delta}A_{(m,n)}^{i}& =(\bar{\delta}\phi _{(m,n)},\bar{\delta}h_{(m,n)},%
\bar{\delta}B_{(m,n)})=2\phi ^{\prime }(\lambda _{(m)}\lambda _{(n)}^{\prime
}-\lambda _{(m)}^{\prime }\lambda _{(n)})(\bar{h},-\bar{\phi},0),  \notag \\
\delta _{a}^{\ast }A_{(m,n)}^{i}& =\left( \delta _{a}^{\ast }\phi
_{(m,n)},\delta _{a}^{\ast }h_{(m,n)},\delta _{a}^{\ast }B_{(m,n)}\right)
=2\phi ^{\prime }(\lambda _{(m)}\lambda _{(n)}^{\prime }-\lambda
_{(m)}^{\prime }\lambda _{(n)})\left( h_{a}^{\ast },-\phi _{a}^{\ast
},0\right) .  \label{var}
\end{align}
From this result, we can see that it is sufficient to consider the variation
of the expressions $\delta _{(l)}(A_{ia}^{\ast }A^{i})$, $\delta _{(l)}(\bar{%
A}_{i}A^{i})$ only with respect to the fields $\phi $ and $h$, since the
variation of $B$ is equal to zero. We have 
\begin{align}
\delta (\delta _{(l)}\phi )& =(\delta \phi )^{\prime }\left( \epsilon
_{(l)}+2\phi ^{\prime }\lambda _{(l)}\right) ,  \notag \\
\delta (\delta _{(l)}h)& =-(\delta h)\epsilon _{(l)}^{\prime }+(\delta
h)^{\prime }\epsilon _{(l)}+2\phi ^{\prime }(\delta \phi )^{\prime
}(B^{\prime }\lambda _{(l)}-B\lambda _{(l)}^{\prime }),  \notag \\
\delta (\delta _{(l)}B)& =-(\delta h)\lambda _{(l)}^{\prime }+2(\delta
h)^{\prime }\lambda _{(l)}.  \label{2var}
\end{align}
Replacing in the above expressions the variations $\delta A^{i}$ with $\bar{%
\delta}A_{(m,n)}^{i}$, $\delta _{a}^{\ast }A_{(m,n)}^{i}$, and substituting
their manifest form, we obtain 
\begin{align*}
& \left. \delta \left( \delta _{(l)}A_{ia}^{\ast }A^{i}\right) \right|
_{\delta A\rightarrow \bar{\delta}A_{(m,n)}}=2\phi _{a}^{\ast }[\bar{h}\phi
^{\prime }(\lambda _{(m)}\lambda _{(n)}^{\prime }-\lambda _{(m)}^{\prime
}\lambda _{(n)})]^{\prime }\left( \epsilon _{(l)}+2\phi ^{\prime }\lambda
_{(l)}\right) \\
& +2h_{a}^{\ast }\Big\{\bar{\phi}\phi ^{\prime }(\lambda _{(m)}\lambda
_{(n)}^{\prime }-\lambda _{(m)}^{\prime }\lambda _{(n)})\epsilon
_{(l)}^{\prime }-[\bar{\phi}\phi ^{\prime }(\lambda _{(m)}\lambda
_{(n)}^{\prime }-\lambda _{(m)}^{\prime }\lambda _{(n)})]^{\prime }\epsilon
_{(l)} \\
& +2\phi ^{\prime }[\bar{h}\phi ^{\prime }(\lambda _{(m)}\lambda
_{(n)}^{\prime }-\lambda _{(m)}^{\prime }\lambda _{(n)})]^{\prime
}(B^{\prime }\lambda _{(l)}-B\lambda _{(l)}^{\prime })\Big\} \\
& +2B_{a}^{\ast }\left\{ \bar{\phi}\phi ^{\prime }(\lambda _{(m)}\lambda
_{(n)}^{\prime }-\lambda _{(m)}^{\prime }\lambda _{(n)})\lambda
_{(l)}^{\prime }-2[\bar{\phi}\phi ^{\prime }(\lambda _{(m)}\lambda
_{(n)}^{\prime }-\lambda _{(m)}^{\prime }\lambda _{(n)})]^{\prime }\lambda
_{(l)}\right\} \,.
\end{align*}
Similarly, we have 
\begin{align*}
& \left. \delta (\delta _{(l)}\bar{A}_{i}A^{i})\right| _{\delta A\rightarrow
\delta ^{\ast }A_{(m,n)}}=2\bar{\phi}[h_{a}^{\ast }\phi ^{\prime }(\lambda
_{(m)}\lambda _{(n)}^{\prime }-\lambda _{(m)}^{\prime }\lambda _{(n)})%
]^{\prime }(\epsilon _{(l)}+2\phi ^{\prime }\lambda _{(l)}) \\
& +2\bar{h}\Big\{\phi _{a}^{\ast }\phi ^{\prime }(\lambda _{(m)}\lambda
_{(n)}^{\prime }-\lambda _{(m)}^{\prime }\lambda _{(n)})\epsilon
_{(l)}^{\prime }-[\phi _{a}^{\ast }\phi ^{\prime }(\lambda _{(m)}\lambda
_{(n)}^{\prime }-\lambda _{(m)}^{\prime }\lambda _{(n)})]^{\prime }\epsilon
_{(l)} \\
& +2\phi ^{\prime }[h_{a}^{\ast }\phi ^{\prime }(\lambda _{(m)}\lambda
_{(n)}^{\prime }-\lambda _{(m)}^{\prime }\lambda _{(n)})]^{\prime
}(B^{\prime }\lambda _{(l)}-B\lambda _{(l)}^{\prime })\Big\} \\
& +2\bar{B}\left\{ \phi _{a}^{\ast }\phi ^{\prime }(\lambda _{(m)}\lambda
_{(n)}^{\prime }-\lambda _{(m)}^{\prime }\lambda _{(n)})\lambda
_{(l)}^{\prime }-2[\phi _{a}^{\ast }\phi ^{\prime }(\lambda _{(m)}\lambda
_{(n)}^{\prime }-\lambda _{(m)}^{\prime }\lambda _{(n)})\lambda
_{(n)})]^{\prime }\lambda _{(l)}\right\}
\end{align*}
and 
\begin{align}
& \left. \delta (\delta _{(l)}\bar{A}_{i}A^{i})\right| _{\delta A\rightarrow 
\bar{\delta}A_{(m,n)}}=2\bar{\phi}[\bar{h}\phi ^{\prime }(\lambda
_{(m)}\lambda _{(n)}^{\prime }-\lambda _{(m)}^{\prime }\lambda _{(n)})%
]^{\prime }(\epsilon _{(l)}+2\phi ^{\prime }\lambda _{(l)})  \notag \\
& +2\bar{h}\Big\{\bar{\phi}\phi ^{\prime }[\lambda _{(m)}\lambda
_{(n)}^{\prime }-\lambda _{(m)}^{\prime }\lambda _{(n)}]\epsilon
_{(l)}^{\prime }-[\bar{\phi}\phi ^{\prime }(\lambda _{(m)}\lambda
_{(n)}^{\prime }-\lambda _{(m)}^{\prime }\lambda _{(n)})]^{\prime }\epsilon
_{(l)}  \notag \\
& +2\phi ^{\prime }[\bar{h}\phi ^{\prime }(\lambda _{(m)}\lambda
_{(n)}^{\prime }-\lambda _{(m)}^{\prime }\lambda _{(n)})]^{\prime
}(B^{\prime }\lambda _{(l)}-B\lambda _{(l)}^{\prime })\Big\}  \notag \\
& +2\bar{B}\left\{ \bar{\phi}\phi ^{\prime }(\lambda _{(m)}\lambda
_{(n)}^{\prime }-\lambda _{(m)}^{\prime }\lambda _{(n)})\lambda
_{(l)}^{\prime }-2[\bar{\phi}\phi ^{\prime }(\lambda _{(m)}\lambda
_{(n)}^{\prime }-\lambda _{(m)}^{\prime }\lambda _{(n)})]^{\prime }\lambda
_{(l)}\right\} .  \label{aa}
\end{align}

Gathering together the contributions (\ref{struct}) corresponding to $F$ and 
$F_{a}$, in (\ref{a1}), (\ref{a2}), we can calculate $S_{3}$ by
differentiating $F$ and $F_{a}$ with respect to Grassmann parameters,
according to (\ref{g_param}), (\ref{s3new}). Using the notation $C^{\alpha
a}=(c^{a},l^{a})$, $B^{\alpha }=(u,v)$ and (\ref{alavan}), we obtain the
expression (\ref{s3}).

{\Large \textbf{Appendix B}}

\setcounter{equation}{0} \renewcommand{\theequation}{B.\arabic{equation}} In
this Appendix, we shall prove that the contributions $S_{1}$, $S_{2}$, $%
S_{3} $ are not sufficient to provide a closed solution to the classical
master equation (\ref{sp(2)_eq}) of the Sp(2)-covariant formalism, where a
solution is sought as an expansion (\ref{series}). To this end, note that
for irreducible theories the contribution $S_{1}$ can be chosen in the form 
\cite{blt}

\begin{equation*}
S_{1}=A_{ia}^{\ast }R_{\alpha }^{\imath }C^{\alpha a}+\bar{A}_{i}R_{\alpha
}^{i}B^{\alpha }-\varepsilon ^{ab}C_{\alpha ab}^{\ast }B^{\alpha }.
\end{equation*}
Then the higher contributions $S_{n+1}$ for $n\geq 1$ can be determined by
iterations: 
\begin{equation}
W^{a}S_{n+1}=F_{n+1}^{a},  \label{s-f}
\end{equation}
where $W^{a}$ is a doublet of differential operators, which in the case $%
\varepsilon (\xi ^{\alpha })=0$ has the form 
\begin{equation*}
W^{a}=S_{0,i}\frac{\delta }{\delta A_{ia}^{\ast }}+A_{ib}^{\ast }R_{\alpha
}^{i}\frac{\delta }{\delta C_{\alpha ab}^{\ast }}+(\bar{A}_{i}R_{\alpha
}^{i}-\varepsilon ^{bc}C_{\alpha bc}^{\ast })\frac{\delta }{\delta B_{\alpha
a}^{\ast }}+\varepsilon ^{ab}B^{\alpha }\frac{\delta _{l}}{\delta C_{\alpha
b}}+\varepsilon ^{ab}\phi _{Aa}^{\ast }\frac{\delta }{\delta \bar{\phi}_{A}}%
\,,
\end{equation*}
and the quantities $F_{n+1}^{a}$ are given by 
\begin{equation*}
F_{n+1}^{a}=-\frac{1}{2}(S_{[n]},S_{[n]})_{n+1}^{a},\;\;\;S_{[n]}=S_{0}+%
\sum_{k=1}^{n}S_{k},
\end{equation*}
with $(\;,\;)_{n+1}^{a}$ being the $(n+1)$-th order of the extended
antibracket in powers of $C^{a\alpha }$, $B^{\alpha }$, 
\begin{equation*}
(F,G)^{a}=\frac{\delta F}{\delta \phi ^{A}}\frac{\delta G}{\delta \phi
_{Aa}^{\ast }}-(-1)^{(\varepsilon (F)+1)(\varepsilon (G)+1)}\frac{\delta G}{%
\delta \phi ^{A}}\frac{\delta F}{\delta \phi _{Aa}^{\ast }}
\end{equation*}
having the obvious property 
\begin{equation*}
(F,G)^{a}=-(-1)^{(\varepsilon (F)+1)(\varepsilon (G)+1)}(G,F)^{a}.
\end{equation*}

Let us assume that the functional $S_{[3]}$, given by 
\begin{equation*}
S=S_{[3]}=S_{0}+S_{1}+S_{2}+S_{3},
\end{equation*}
is a close solution of CME. Then the quantities $F_{n+1}^{a}$ for $n\geq 3$
must vanish identically: 
\begin{equation*}
S_{n+1}=0\Rightarrow W^{a}S_{n+1}=F_{n+1}^{a}=0,\;\;\;n\geq 3.
\end{equation*}
Considering all possible $F_{n+1}^{a}$, $n\geq 3$, we have 
\begin{align*}
F_{4}^{a}& =-\frac{1}{2}%
(S_{[3]},S_{[3]})_{4}^{a}=-(S_{1},S_{3})_{4}^{a}-(S_{2},S_{3})_{4}^{a}-\frac{%
1}{2}(S_{2},S_{2})_{4}^{a}\,, \\
F_{5}^{a}& =-\frac{1}{2}(S_{[4]},S_{[4]})_{5}^{a}=-\frac{1}{2}%
(S_{[3]},S_{[3]})_{5}^{a}=-(S_{2},S_{3})_{5}^{a}-\frac{1}{2}%
(S_{3},S_{3})_{5}^{a}\,, \\
F_{6}^{a}& =-\frac{1}{2}(S_{[5]},S_{[5]})_{6}^{a}=-\frac{1}{2}%
(S_{[3]},S_{[3]})_{6}^{a}=-\frac{1}{2}(S_{3},S_{3})_{6}^{a}.
\end{align*}
Note that $F_{n+1}^{a}\equiv 0$, $n\geq 6$. Thus, the quantity $F_{6}^{a}$
has the simplest structure, which involves only the contribution $S_{3}$. In
what follows, we shall check if the condition $(S_{3},S_{3})_{6}^{a}=0$ is
fulfilled in the model of $W_{3}$-gravity, which is necessary for $S_{[3]}$
to be a closed solution of CME.

Let us consider the expression (\ref{s3_sp(2)}) for $S_{3}$. Then $%
(S_{3},S_{3})_{6}^{a}$, given by 
\begin{equation*}
\frac{1}{2}(S_{3},S_{3})_{6}^{a}=\frac{\delta S_{3}}{\delta A^{i}}\frac{%
\delta S_{3}}{\delta A_{ia}^{\ast }}\,,
\end{equation*}
decomposes into the following orders in antifields: 
\begin{equation*}
\bar{A}(2\bar{C}-B^{\ast }),\;\;\;(\bar{A})^{2}A^{\ast },\;\;\;(\bar{A}%
)^{3}\,,
\end{equation*}
namely, 
\begin{equation*}
\frac{1}{2}(S_{3},S_{3})_{6}^{a}=\frac{1}{(12)^{2}}D_{1}^{a}-\frac{1}{%
(12)^{2}}D_{2}^{a}+\frac{1}{24}D_{3}^{a},\;\;\;D_{n}^{a}\sim (\bar{A})^{n},
\end{equation*}
where 
\begin{align*}
D_{1}^{a}& =\bar{A}_{m}(2\bar{C}_{\nu d}-B_{\nu d}^{\ast })(F_{\rho \lambda
}^{\nu }F_{\delta \sigma }^{\lambda }+2F_{\rho \delta ,k}^{\nu }R_{\sigma
}^{k})_{,l}(2R_{\beta ,n}^{l}M_{\alpha \gamma }^{mn}+4R_{\beta
,n}^{m}M_{\alpha \gamma }^{ln}- \\
& -M_{\beta \lambda }^{lm}F_{\alpha \gamma }^{\lambda }-2M_{\beta \alpha
,n}^{lm}R_{\gamma }^{n})C^{\gamma c}C^{\beta b}C^{\alpha a}C^{\sigma
q}C^{\rho p}C^{\delta d}\varepsilon _{bc}\varepsilon _{pq}\,, \\
D_{2}^{a}& =A_{id}^{\ast }\bar{A}_{j}\bar{A}_{m}(2R_{\rho ,k}^{i}M_{\delta
\sigma }^{jk}+4R_{\rho ,k}^{j}M_{\delta \sigma }^{ik}-M_{\rho \lambda
}^{ij}F_{\delta \sigma }^{\lambda }-2M_{\rho \delta ,k}^{ij}R_{\sigma
}^{k})_{,l}(2R_{\beta ,n}^{l}M_{\alpha \gamma }^{mn}+ \\
& +4R_{\beta ,n}^{m}M_{\alpha \gamma }^{ln}-M_{\beta \lambda }^{lm}F_{\alpha
\gamma }^{\lambda }-2M_{\beta \alpha ,n}^{lm}R_{\gamma }^{n})C^{\gamma
c}C^{\beta b}C^{\alpha a}C^{\sigma q}C^{\rho p}C^{\delta d}\varepsilon
_{bc}\varepsilon _{pq}\,, \\
D_{3}^{a}& =\bar{A}_{i}\bar{A}_{j}\bar{A}_{m}(R_{\rho ,k}^{i}M_{\delta
\sigma }^{jk})_{,l}(2R_{\beta ,n}^{l}M_{\alpha \gamma }^{mn}+4R_{\beta
,n}^{m}M_{\alpha \gamma }^{ln}-M_{\beta \lambda }^{lm}F_{\alpha \gamma
}^{\lambda } \\
& -2M_{\beta \alpha ,n}^{lm}R_{\gamma }^{n})C^{\gamma c}C^{\beta b}C^{\alpha
a}C^{\sigma q}C^{\rho p}B^{\delta }\varepsilon _{bc}\varepsilon _{pq}\,.
\end{align*}

We can see that the quantity $D_{3}^{a}$, given by the order $(\bar{A})^{3}$%
, has the simplest form. Therefore, in what follows we shall investigate the
question whether $D_{3}^{a}$ is equal to zero in the model of $W_{3}$%
-gravity.

Before taking into account the manifest form of the gauge algebra of the
given model, let us simplify the consideration by rewriting $D_{3}^{a}$ in
the following manner: 
\begin{equation*}
D_{3}^{a}=\varepsilon _{bc}\varepsilon _{pq}\frac{\partial _{r}}{\partial
\mu _{p(5)}}\frac{\partial _{r}}{\partial \mu _{q(4)}}\frac{\partial _{r}}{%
\partial \mu _{a(3)}}\frac{\partial _{r}}{\partial \mu _{b(2)}}\frac{%
\partial _{r}}{\partial \mu _{c(1)}}D.
\end{equation*}
where $D$ is given by 
\begin{align*}
D& =\bar{A}_{j}[(\delta _{(5)}A^{j})_{,k}(\bar{A}_{i}\bar{\xi}%
_{(4,0)}^{ik})]_{,l}[2(\delta _{(2)}A^{l})_{,n}(\bar{A}_{m}\bar{\xi}%
_{(1,3)}^{mn}) \\
& -4(\delta _{(2)}\bar{A}_{m}A^{m})_{,n}\bar{\xi}_{(1,3)}^{nl}+\bar{A}_{m}%
\bar{\xi}_{((1,3),2)}^{ml}+2\delta _{(1)}(\bar{A}_{m}\bar{\xi}%
_{(3,2)}^{ml})],
\end{align*}
Here, $\delta _{(n)}$ stand for gauge variations with parameters $\xi
_{(n)}^{\alpha }$, $n=0,1,...,5$, defined by (\ref{g_param}), while the
quantities $\bar{\xi}_{(m,n)}^{ij}$, $\bar{\xi}_{((m,n),l)}^{ij}$ are given
by (\ref{param4}), (\ref{param5}). Using this notation, we can rewrite the
quantity $D$ in a compact form: 
\begin{equation*}
D=\bar{A}_{i}\frac{\delta }{\delta A^{j}}\left( \delta \left. (\delta
_{(5)}A^{i})\right| _{\delta A\longrightarrow \bar{\delta}A_{(4,0)}}\right)
\delta A_{(1,3,2)}^{j}=\left. \delta \left( \delta \left. (\delta _{(5)}\bar{%
A}_{i}A^{i})\right| _{\delta A\longrightarrow \bar{\delta}A_{(4,0)}}\right)
\right| _{\delta A\longrightarrow \delta A_{(1,3,2)}}\,,
\end{equation*}
where the variations $\delta $ are understood as usual variations with
respect to the fields $A^{i}$. To obtain $D$, one has to perform two
successive variations of the quantity $\delta _{(5)}(\bar{A}_{i}A^{i})$,
replacing the corresponding variations $\delta A^{i}$ in the resulting
expressions $\delta (\delta _{(5)}\bar{A}_{i}A^{i})$ and $\delta (\delta
(\delta _{(5)}\bar{A}_{i}A^{i}))$ by the quantities $\bar{\delta}%
A_{(4,0)}^{i}$ and $\delta A_{(1,3,2)}^{i}$%
\begin{align*}
& \bar{\delta}A_{(4,0)}^{i}=\bar{A}_{j}\bar{\xi}_{(4,0)}^{ji},\;\;\delta
A_{(1,3,2)}^{i}=2(\delta _{(2)}A^{i})_{,k}(\bar{A}_{j}\bar{\xi}_{(1,3)}^{jk})
\\
& \,-4(\delta _{(2)}\bar{A}_{k}A^{k})_{,j}\bar{\xi}_{(1,3)}^{ji}+\bar{A}_{j}%
\bar{\xi}_{((1,3),2)}^{ji}+2\delta _{(1)}(\bar{A}_{j}\bar{\xi}_{(3,2)}^{ji}).
\end{align*}

To calculate the quantity $D$ explicitly, we remind that in the model of $%
W_{3}$-gravity we have $A^{i}=(\phi,h,B)$, $\xi^{\alpha}=(\epsilon,\lambda)$%
. The gauge transformations $\delta A^{i}=R_{\alpha}^{i}\xi^{\alpha}$ are
given by (\ref{g_tr}) and the (non-vanishing) quantities $\tilde{\xi}%
_{(m,n)}^{\alpha}$, $\bar{\xi}_{(m,n)}^{ij}$, corresponding to subsequent
gauge transformations with the parameters $\xi_{(m)}^{\alpha}$ and $\xi
_{(n)}^{\alpha}$, have the form (\ref{param7}).

The quantity $\delta \left. (\delta _{(5)}\bar{A}_{i}A^{i})\right| _{\delta
A\longrightarrow \bar{\delta}A_{(4,0)}}$ is given by (\ref{aa}). Taking a
variation of this expression, we find 
\begin{align}
& \delta \left( \delta \left. (\delta _{(5)}\bar{A}_{i}A^{i})\right|
_{\delta A\longrightarrow \bar{\delta}A_{(4,0)}}\right) =2\bar{\phi}[\bar{h}%
(\delta \phi )^{\prime }(\lambda _{(4)}\lambda _{(0)}^{\prime }-\lambda
_{(4)}^{\prime }\lambda _{(0)})]^{\prime }(\epsilon _{(5)}+2\phi ^{\prime
}\lambda _{(5)})  \notag \\
& +2\bar{\phi}[\bar{h}(\delta \phi )^{\prime }(\lambda _{(4)}\lambda
_{(0)}^{\prime }-\lambda _{(4)}^{\prime }\lambda _{(0)})]^{\prime }\delta
\phi ^{\prime }\lambda _{(5)}+2\bar{h}\Big\{\bar{\phi}(\delta \phi )^{\prime
}(\lambda _{(4)}\lambda _{(0)}^{\prime }-\lambda _{(4)}^{\prime }\lambda
_{(0)})\epsilon _{(5)}^{\prime }  \notag \\
& -[\bar{\phi}(\delta \phi )^{\prime }(\lambda _{(4)}\lambda _{(0)}^{\prime
}-\lambda _{(4)}^{\prime }\lambda _{(0)})]^{\prime }\epsilon _{(5)}+4(\delta
\phi )^{\prime }[\bar{h}\phi ^{\prime }(\lambda _{(4)}\lambda _{(0)}^{\prime
}-\lambda _{(4)}^{\prime }\lambda _{(0)})]^{\prime }(B^{\prime }\lambda
_{(5)}-B\lambda _{(5)}^{\prime })  \notag \\
& +4\phi ^{\prime }[\bar{h}(\delta \phi )^{\prime }(\lambda _{(4)}\lambda
_{(0)}^{\prime }-\lambda _{(4)}^{\prime }\lambda _{(0)})]^{\prime
}(B^{\prime }\lambda _{(5)}-B\lambda _{(5)}^{\prime })  \notag \\
& +4\phi ^{\prime }[\bar{h}\phi ^{\prime }(\lambda _{(4)}\lambda
_{(0)}^{\prime }-\lambda _{(4)}^{\prime }\lambda _{(0)})]^{\prime }[(\delta
B)^{\prime }\lambda _{(5)}-(\delta B)\lambda _{(5)}^{\prime }]\Big\}  \notag
\\
& +2\bar{B}\Big\{\bar{\phi}(\delta \phi )^{\prime }(\lambda _{(4)}\lambda
_{(0)}^{\prime }-\lambda _{(4)}^{\prime }\lambda _{(0)})\lambda
_{(5)}^{\prime }-2[\bar{\phi}(\delta \phi )^{\prime }(\lambda _{(4)}\lambda
_{(0)}^{\prime }-\lambda _{(4)}^{\prime }\lambda _{(0)})]^{\prime }\lambda
_{(5)}\Big\}.  \label{delta3}
\end{align}

Notice that the above expression does not contain any variations with
respect to the field $h$. Therefore, in what follows it is sufficient to
find only two components of the variation $\delta A_{(1,3,2)}^{i}$ needed to
calculate the quantity $D$, namely, $\delta\phi_{(1,3,2)}$ and $\delta
B_{(1,3,2)}$. With this in mind, we observe that in order to construct $%
\delta A_{(1,3,2)}^{i}$ we need to determine the following set of
quantities: 
\begin{equation*}
(\delta_{(2)}A^{i})_{,k}(\bar{A}_{j}\bar{\xi}_{(1,3)}^{jk})\,,\;\;\;(\delta
_{(2)}\bar{A}_{k}A^{k})_{,j}\bar{\xi}_{(1,3)}^{ji}\,,\;\;\;\bar{A}_{j}\bar
{\xi}_{((1,3),2)}^{ji}\,,\;\;\;\delta_{(1)}(\bar{A}_{j}\bar{\xi}%
_{(3,2)}^{ji})\,,
\end{equation*}
where $i=(\phi,B)$, $j,k=(\phi,h,B)$. These quantities can be easily
calculated using (\ref{g_tr}), (\ref{param8}), (\ref{xi3}), (\ref{varxi}).
We have

\begin{align*}
& \delta \phi _{(1,3,2)}=2(\delta _{(2)}\phi )_{,k}(\bar{A}_{j}\bar{\xi}%
_{(1,3)}^{jk})-4(\delta _{(2)}\bar{A}_{k}A^{k})_{,j}\bar{\xi}_{(1,3)}^{j\phi
}+\bar{A}_{j}\bar{\xi}_{((1,3),2)}^{j\phi }+2\delta _{(1)}(\bar{A}_{j}\bar{%
\xi}_{(3,2)}^{j\phi }) \\
& =4[\bar{h}\phi ^{\prime }(\lambda _{(1)}\lambda _{(3)}^{\prime }-\lambda
_{(1)}^{\prime }\lambda _{(3)})]^{\prime }(\epsilon _{(2)}+2\phi ^{\prime
}\lambda _{(2)})-8\phi ^{\prime }(\lambda _{(1)}\lambda _{(3)}^{\prime
}-\lambda _{(1)}^{\prime }\lambda _{(3)}) \\
& \times \lbrack \bar{h}\epsilon _{(2)}^{\prime }+\bar{B}\lambda
_{(2)}^{\prime }+(\bar{h}\epsilon _{(2)}+2\bar{B}\lambda _{(2)})^{\prime
}]^{\prime }+2\bar{h}\phi ^{\prime }(\epsilon _{(1)}\lambda _{(3)}^{\prime
}-2\epsilon _{(1)}^{\prime }\lambda _{(3)}-\lambda _{(1)}^{\prime }\epsilon
_{(3)}+2\lambda _{(1)}\epsilon _{(3)}^{\prime })\lambda _{(2)}^{\prime } \\
& -2\bar{h}\phi ^{\prime }(\epsilon _{(1)}\lambda _{(3)}^{\prime }-2\epsilon
_{(1)}^{\prime }\lambda _{(3)}-\lambda _{(1)}^{\prime }\epsilon
_{(3)}+2\lambda _{(1)}\epsilon _{(3)}^{\prime })^{\prime }\lambda _{(2)}+4%
\bar{h}(\phi ^{\prime }\epsilon _{(1)}+\phi ^{\prime \,2}\lambda
_{(1)})^{\prime }(\lambda _{(3)}\lambda _{(2)}^{\prime }-\lambda
_{(3)}^{\prime }\lambda _{(2)}), \\
& \delta B_{(1,3,2)}=2(\delta _{(2)}B)_{,k}(\bar{A}_{j}\bar{\xi}%
_{(1,3)}^{jk})=4\bar{\phi}\phi ^{\prime }(\lambda _{(1)}\lambda
_{(3)}^{\prime }-\lambda _{(1)}^{\prime }\lambda _{(3)})\lambda
_{(2)}^{\prime }-8[\bar{\phi}\phi ^{\prime }(\lambda _{(1)}\lambda
_{(3)}^{\prime }-\lambda _{(1)}^{\prime }\lambda _{(3)})]^{\prime }\lambda
_{(2)}.
\end{align*}

Now, we are able to calculate the quantity $D$, obtained by substituting the
above variations $\delta \phi _{(1,3,2)}$ and $\delta B_{(1,3,2)}$ into (\ref
{delta3}). With this in mind, to simplify the consideration, we will first
analyze the general structure of $D$ as regards its expansion in antifields.
We can write symbolically 
\begin{equation*}
D=\left. (\bar{\phi}\bar{h}+(\bar{h})^{2}+\bar{B}\bar{\phi})\delta \phi
\right| _{\delta \phi =\bar{h}+\bar{B}}+\left. (\bar{h})^{2}\delta B\right|
_{\delta B=\bar{\phi}}\,,
\end{equation*}
which implies that the expression for $D$ decomposes into the following
orders in antifields: 
\begin{equation*}
\bar{\phi}\bar{h}\bar{B},\;\;\;\bar{\phi}(\bar{h})^{2},\;\;\;\bar{\phi}(\bar{%
B})^{2},\;\;\;(\bar{h})^{2}\bar{B},\;\;\;\bar{h}^{3}.
\end{equation*}

From the manifest form of $\delta\phi_{(1,3,2)}$, $\delta B_{(1,3,2)}$ and $%
\delta\left( \delta\left. (\delta_{(5)}\bar{A}_{i}A^{i})\right| _{\delta
A\longrightarrow\bar{\delta}A_{(4,0)}}\right) $, one can observe that the
contribution $\bar{\phi}(\bar{B})^{2}$ has the simplest form, which we shall
denote by $D_{\bar{\phi}\bar{B}^{2}}$. This contribution is given by 
\begin{equation*}
D_{\bar{\phi}\bar{B}^{2}}=\left. D\right| _{\bar{h}=0},
\end{equation*}
and therefore to calculate $D_{\bar{\phi}\bar{B}^{2}}$ it is sufficient to
consider the limits 
\begin{equation*}
\left. \delta\left( \delta\left. (\delta_{(5)}\bar{A}_{i}A^{i})\right|
_{\delta A\longrightarrow\bar{\delta}A_{(4,0)}}\right) \right| _{\bar{h}%
=0}\,,\;\;\left. \delta\phi_{(1,3,2)}\right| _{\bar{h}=0}\,,\;\;\left.
\delta B_{(1,3,2)}\right| _{\bar{h}=0}.
\end{equation*}
We have

\begin{equation*}
D_{\bar{\phi}\bar{B}^{2}}=\left. \delta \left( \delta \left. (\delta _{(5)}%
\bar{A}_{i}A^{i})\right| _{\delta A\longrightarrow \bar{\delta}%
A_{(4,0)}}\right) \right| _{\delta A\longrightarrow \delta
A_{(1,3,2)}},\;\;\;\bar{h}=0,
\end{equation*}
with 
\begin{gather*}
\left. \delta \left( \delta \left. (\delta _{(5)}\bar{A}_{i}A^{i})\right|
_{\delta A\longrightarrow \bar{\delta}A_{(4,0)}}\right) \right| _{\bar{h}%
=0}=-2(\delta \phi )[\bar{\phi}(\lambda _{(4)}\lambda _{(0)}^{\prime
}-\lambda _{(4)}^{\prime }\lambda _{(0)})(3\bar{B}\lambda _{(5)}^{\prime }+2%
\bar{B}^{\prime }\lambda _{(5)})]^{\prime }, \\
\delta \phi =\left. \delta \phi _{(1,3,2)}\right| _{\bar{h}=0}=-8\phi
^{\prime }(\lambda _{(1)}\lambda _{(3)}^{\prime }-\lambda _{(1)}^{\prime
}\lambda _{(3)})(3\bar{B}\lambda _{(2)}^{\prime }+2\bar{B}^{\prime }\lambda
_{(2)})^{\prime },
\end{gather*}
where we have used integration by parts in the expression containing $\bar{A}%
_{i}A^{i}$. Thus, we obtain

\begin{equation*}
D_{\bar{\phi}\bar{B}^{2}}=16\phi ^{\prime }(\lambda _{(1)}\lambda
_{(3)}^{\prime }-\lambda _{(1)}^{\prime }\lambda _{(3)})(3\bar{B}\lambda
_{(2)}^{\prime }+2\bar{B}^{\prime }\lambda _{(2)})^{\prime }[\bar{\phi}%
(\lambda _{(4)}\lambda _{(0)}^{\prime }-\lambda _{(4)}^{\prime }\lambda
_{(0)})(3\bar{B}\lambda _{(5)}^{\prime }+2\bar{B}^{\prime }\lambda
_{(5)})]^{\prime }.
\end{equation*}

The contribution $D_{\bar{\phi}\bar{B}^{2}}$ to $D$ determines the
corresponding order in antifields in the quantity $D_{3}^{a}$, which
represents the order $(\bar{A})^{3}$ in $F_{6}^{a}$. Since $D_{\bar{\phi}%
\bar{B}^{2}}$ is related to $D$ by the limit $\bar{h}=0$, we have 
\begin{equation*}
\left. D_{3}^{a}\right| _{\bar{h}=0}=\varepsilon _{bc}\varepsilon _{pq}\frac{%
\partial _{r}}{\partial \mu _{p(5)}}\frac{\partial _{r}}{\partial \mu _{q(4)}%
}\frac{\partial _{r}}{\partial \mu _{a(3)}}\frac{\partial _{r}}{\partial \mu
_{b(2)}}\frac{\partial _{r}}{\partial \mu _{c(1)}}D_{\bar{\phi}\bar{B}%
^{2}}\,,
\end{equation*}
with $\lambda _{(0)}=v,\;\lambda _{(n)}=l^{a}\mu _{a(n)}$.

Taking derivatives with respect to the Grassmann parameters, and using
integration by parts in the resultant expression, we obtain

\begin{equation*}
\left. D_{3}^{a}\right| _{\bar{h}=0}=16\bar{\phi}\{\phi ^{\prime
}[l^{b}(l^{a})^{\prime }-(l^{b})^{\prime }l^{a}][3\bar{B}(l^{c})^{\prime }+2%
\bar{B}^{\prime }l^{c}]^{\prime }\}^{\prime }[l^{p}v^{\prime
}-(l^{p})^{\prime }v][3\bar{B}(l^{q})^{\prime }+2\bar{B}^{\prime
}l^{q}]\varepsilon _{bc}\varepsilon _{pq}.
\end{equation*}

One can show that $\left. D_{3}^{a}\right| _{\bar{h}=0}$ does not vanish
identically. Since this fact is not evident in the Sp(2)-covariant form, we
shall write the dummy indices $b$, $c$, $p$, $q$ manifestly in terms of the
values $1$, $2$, denoting $l^{1}=l$, $l^{2}=\bar{l}$, and taking into
account $\varepsilon _{12}=-1$. Let us also fix the free index as $a=1$.

Using the cancellation of some terms containing squares of $l$, $\,\bar{l}$, 
$l^{\prime }$, we can represent $\left. D_{3}^{1}\right| _{\bar{h}=0}$ as
follows: 
\begin{equation*}
\left. D_{3}^{1}\right| _{\bar{h}=0}=-16\bar{\phi}\left\{ (\bar{l}%
\,l^{\prime }-\bar{l}^{\prime }\,l)[\phi ^{\prime }(3\bar{B}l^{\prime }+2%
\bar{B}^{\prime }l)^{\prime }]^{\prime }+(\bar{l}\,l^{\,\prime \prime }-\bar{%
l}^{\,\prime \prime }\,l)\phi ^{\prime }(3\bar{B}l^{\prime }+2\bar{B}%
^{\prime }l)^{\prime }\right\} D^{12},
\end{equation*}
where 
\begin{equation*}
D^{12}\equiv (l\,\bar{l}^{\prime }+l^{\prime }\,\bar{l})(3\bar{B}v^{\prime
}-2\bar{B}^{\prime }v)-6l^{\prime }\,\bar{l}^{\prime }\bar{B}v.
\end{equation*}
From this representation of $\left. D_{3}^{1}\right| _{\bar{h}=0}$ it is not
yet evident if this quantity vanishes or not.

To examine the structure of $\left. D_{3}^{1}\right| _{\bar{h}=0}$ in more
detail, let us represent $\left. D_{3}^{1}\right| _{\bar{h}=0}$ in the form 
\begin{equation*}
\left. D_{3}^{1}\right| _{\bar{h}=0}=-16\bar{\phi}(A+B),
\end{equation*}
where 
\begin{eqnarray*}
A &\equiv &(\bar{l}\,l^{\prime }-\bar{l}^{\prime }\,l)[\phi ^{\prime }(3\bar{%
B}l^{\prime }+2\bar{B}^{\prime }l)^{\prime }]^{\prime }D^{12}, \\
B &\equiv &(\bar{l}\,l^{\prime \prime }-\bar{l}^{\prime \prime }l)\phi
^{\prime }(3\bar{B}l^{\prime }+2\bar{B}^{\prime }l)^{\prime }D^{12}.
\end{eqnarray*}
Note that $\left. D_{3}^{1}\right| _{\bar{h}=0}$ vanishes if and only if $%
A+B=0$ because $\bar{\phi}$ is an arbitrary field.

We can see that $A=0$. Indeed, taking into account the manifest form of $%
D^{12}$, we have

\begin{equation*}
A\sim (\bar{l}\,l^{\prime }-\bar{l}^{\prime }l)(l\,\bar{l}^{\prime
}+l^{\prime }\,\bar{l})=(l\,\bar{l}^{\prime })^{2}-(l^{\prime }\,\bar{l}%
)^{2}\equiv 0.
\end{equation*}

Let us consider the quantity $B$. Using the manifest form of $D^{12}$ and
taking into account the cancellation of terms containing squares of $l$, $%
\bar{l}$, $l^{\prime }$, $l^{\prime \prime }$, we obtain 
\begin{equation*}
B=\phi ^{\prime }l\,l^{\prime }\,l^{\prime \prime }[(4v\bar{B}\bar{B}%
^{\prime \prime }-10v\bar{B}^{\prime \,2}+15v^{\prime }\bar{B}\bar{B}%
^{\prime })\bar{l}\,\bar{l}^{\prime }-(6v\bar{B}^{\prime }-9v^{\prime }\bar{B%
})\bar{B}\bar{l}\,\bar{l}^{\prime \prime }-6v\bar{B}^{2}\bar{l}^{\prime }\,%
\bar{l}^{\prime \prime }].
\end{equation*}
This expression does not vanish identically, and therefore $A+B\neq 0$.

Note that since $(S_{3},S_{3})_{6}^{a}$ makes a non-vanishing contribution
to the r.h.s. of the equation $W^{a}S_{6}=F_{6}^{a}$ in (\ref{s-f}) the
contribution $S_{6}$ may be non-vanishing.

\end{document}